\documentclass[bibyear]{aa} 

\usepackage{graphicx}
\usepackage{natbib,twoopt}
\usepackage{color}
\usepackage{lscape}
\usepackage{longtable}
\usepackage{times}
\usepackage{nccmath}
\usepackage{float}

\begin{document}

   \title{\textcolor{blue}{Investigation of dust attenuation  and star formation activity  in galaxies hosting GRBs}}

   \author{D. Corre       \inst{1}
          \and V. Buat    \inst{1}
          \and S. Basa    \inst{1}
          \and S. Boissier \inst{1}
          \and J. Japelj  \inst{2}
          \and J. Palmerio \inst{3}
          \and R. Salvaterra   \inst{4}
          \and S. D. Vergani  \inst{5}
          \and T. Zafar   \inst{6} 
          }

   \institute{
   Aix Marseille Univ, CNRS, CNES, LAM, Marseille, France
   \\   
   \email{david.corre@lam.fr, veronique.buat@lam.fr}
   \and
   Anton Pannekoek Institute for Astronomy, University of Amsterdam, Science Park 904, 1098 XH Amsterdam, The Netherlands 
   \and
   Institut d’Astrophysique de Paris, Université Paris 6-CNRS, UMR7095, 98bis Boulevard Arago, F-75014 Paris, France
   \and
   INAF–IASF Milano, via E. Bassini 15, 20133 Milano, Italy  
   \and
   GEPI, Observatoire de Paris, PSL University, CNRS,  5 Place Jules Janssen, 92190 Meudon, France 
   \and
   Australian Astronomical Observatory, PO Box 915, North Ryde, NSW 1670, Australia
    }


 
  \abstract
   {The gamma-ray bursts hosts (GRBHs) are excellent targets to study the extinction properties of dust and its effects on the global emission of distant galaxies. The dust extinction curve is measured along the GRB afterglow line of sight and the analysis of the spectral energy distribution (SED) of the host galaxy gives access to the global dust attenuation of the stellar light.}
   { In this pilot study we gather information on dust extinction in GRBHs to compare  the properties of the extinction curve to those of the dust obscuration affecting the total stellar light of the host galaxy.  
   Assuming the extinction curve to be representative of the dust properties, we aim to investigate which dust-stars geometries and  local dust distribution in the inter stellar medium (ISM) can reproduce the observed attenuation curve. }
   {We selected a sample of 30 GRBs for which the extinction curve along the GRB afterglow line-of-sight (l.o.s) is measured in the rest-frame ultraviolet (UV) up to optical and  we analysed the properties of the extinction curve as a function of the host galaxy properties. From these 30 GRBs, we selected seven GRBHs with a good rest-frame UV to near infra-red (NIR) spectral coverage for the host. The attenuation curve was derived by fitting the SEDs of the GRBH sample with the CIGALE SED fitting code. Different star formation histories (SFH) were studied to recover the star formation rates (SFR) derived using H$_{\alpha}$ luminosities. Implications for the dust-stars geometries in the ISM are inferred by a comparison with radiative transfer simulations.}
  {The most extinguished GRBs are preferentially found in the more massive hosts and the UV bump is preferentially found in the most extinguished GRB l.o.s. Five out of seven hosts are best fitted with a recent burst of star formation, leading to lower stellar mass estimates than previously found. The average attenuation in the host galaxies is about 70\% of the amount of extinction along the GRB l.o.s. We find a great variety in the derived attenuation curves of GRBHs, the UV slope can be similar, flatter or even steeper than the extinction curve slope. Half of the attenuation curves are consistent with the Calzetti attenuation law and there is evidence of a UV bump in only one GRBH. We find that the flatter (steeper) attenuation curves are found in galaxies with the highest (lowest) SFR and stellar masses. The comparison of our results with radiative transfer simulations leads to a uniform distribution of dust and stars in a very clumpy ISM for half the GRBHs and various dust-stars geometries for the second half of the sample.}
   {}


   \maketitle
%

\section{Introduction}
Stellar light in galaxies is absorbed and scattered in the interstellar medium by dust grains. Dust effects are particularly strong in the ultraviolet (UV), where photons are preferentially absorbed and scattered by interstellar dust. Moreover the UV light is emitted by young stars and is a tracer of the recent star formation. It is thus of the utmost importance to well quantify the interaction between the dust and the stellar light in the UV to study the dust properties and to measure accurately the star formation activity inside a galaxy. \\
Dust attenuation for extended objects should be distinguished from the extinction for point sources. In the case of a point source, the extinction quantifies how many photons are absorbed and scattered out of the line of sight (l.o.s) and is thus directly linked to dust properties. For extended objects, such as galaxies, dust and stars are mixed and the attenuation does not only quantify how many photons are absorbed and scattered by dust out of the l.o.s but also how many photons are scattered into the l.o.s. This means that it does not only depend on dust properties but also on the dust-stars geometry in the ISM.  \\
The extinction curve can be directly measured along sightlines where the spectrum of individual stars can be obtained, which due to resolution is only the case in the Local Group so far. Using this technique, extinction curves were obtained in four galaxies where stars can still be resolved: the Milky Way \citep[MW;][]{Witt84,Fitzpatrick90,Fitzpatrick07}, the Small Magellanic Cloud \citep[SMC;][]{Prevot84,Gordon98,Gordon03}, the Large Magellanic Cloud \citep[LMC;][]{Clayton85,Fitzpatrick85,Misselt99}, and M31 (\citealt{Bianchi96,Clayton15}).  \\
Beyond the Local Group, where individual stars cannot be resolved, several techniques have been developed to derive the dust extinction in extragalactic sources, provided that the intrinsic SEDs of these sources are well characterised (see a more detailed review in \citealt{Draine03}). Most of these techniques assume the sources to be point-like which is not the case and therefore geometrical effects are included in the extinction curve. The importance of these effects are discussed by \cite{Kruegel09} and \cite{Scicluna15}. For example, the dust extinction curve can be derived from QSO sightlines by comparing composite spectra of reddened and un-reddened quasars (\citealt{Richards03,Hopkins04,Zafar15}) and it is found that featureless SMC-like extinction curves are the most common along QSO sightlines.

The attenuation curve of a whole galaxy can be derived from observations using UV-optical spectra or multi-band photometric data. For local UV-bright starburst galaxies, \cite{Calzetti94, Calzetti00} derive a mean attenuation curve characterised by a flatter slope than for mean local extinction curves and by a lack of the 2175 A absorption feature (UV bump). 
After Calzetti's work the topic was reinvestigated at low and high redshift \citep[e.g.][]{Buat10,Buat11,Buat12,Wild11,Kriek13,Reddy15,Reddy16, Salmon16, Battisti16,Battisti17} leading to a non-universality of the dust attenuation law. 
Using a similar method as the one of \cite{Calzetti00} (hereafter C00) on a larger sample of about 10000 local (z $\lesssim$ 0.1) star forming galaxies, \cite{Battisti16,Battisti17} derive a mean attenuation curve similar to C00 one though a bit shallower in the UV.  \cite{Reddy15} also applied a similar method to the C00 one to estimate the attenuation curve of 224 star forming galaxies at z$\sim$2. They find that the mean attenuation curve shape is again similar to the one of C00 in the UV but with a lower normalisation and similar to the mean SMC extinction curve at larger wavelength. Based on the NOAO Extremely Wide Field Infrared Mosaic (NEWFIRM) Medium-Band Survey, \cite{Kriek13} constrain the dust attenuation of 0.5 < z < 2 galaxies. They find that a MW and Calzetti law provide poor fits at UV wavelengths, and that steeper laws have stronger UV bumps together with more active galaxies having flatter attenuation curves and weaker UV bumps. Using galaxies from the Cosmic Assembly Near-infrared Deep Extragalactic Legacy Survey (CANDELS) at z$\sim$1.5-3, \cite{Salmon16} find a correlation between the colour excess E(B-V) and the slope of the attenuation law: galaxies with high colour excess have shallower attenuation curves and those with low colour excess have steeper attenuation curves.\\

The attenuation curve of galaxies is also studied using radiative transfer models with various dust physical properties and geometrical distribution of stars and dust. In general, shallower attenuation curves correspond to larger optical depths and mixed  star-dust geometries (e.g. \citealt{Witt00, Pierini04,Tuffs04,Panuzzo07,Chevallard13}).  
Applying radiative transfer models to M31 and M51, \cite{Viaene17} and \cite{DeLooze14} respectively derived very steep attenuation curves with the presence of a UV bump. 
Most of the numerical simulations using radiative transfer models use the extinction law of local galaxies to characterise the dust properties and compute the attenuation curve one would observe for a given dust-stars geometry (e.g. \citealt{Witt00,Pierini04,Seon16}).\\

Gamma-ray bursts (GRBs) with their featureless power-law like spectrum shape of their afterglows are excellent probes of dust extinction in their host galaxies (\citealt{Zafar11}; \citealt{Schady12}; \citealt{Covino13}; \citealt{Japelj15}; Zafar et al. 2018 (submitted)). Thanks to their extremely high luminosity (up to $\sim$ 10$^{53}$ erg.s$^{-1}$) such studies can be extended to very high redshift. In this work, we take advantage of the fact that both the extinction and attenuation curves can be measured for the same galaxy. The observed difference between attenuation and extinction curves contains information about the GRBH ISM structure. The extinction curves measured along the GRB l.o.s are taken from the literature. The global dust attenuation in the GRBH galaxies is obtained by deriving the attenuation curve from SED fitting. Assuming GRB extinction curves are representative of the dust properties of their hosts, we compared our results to radiative transfer simulations in order to put some constraint on the dust-stars geometry and local distribution of dust in the ISM. As a by-product of the SED fitting we also derive robust stellar masses and SFR for the GRBHs.
\\
The paper is organised as follows: in Section 2, we explain the selection of the GRB extinction curves and later the selection of the GRB host galaxies. In Section 3 we will search for correlations between the GRB l.o.s and the host galaxy properties. In Section 4, we derive the attenuation curves of the selected host galaxies using the CIGALE SED fitting algorithm as well as the stellar mass and SFR of the galaxies. In Section 5, the extinction curves are compared to the derived attenuation curves and we compare our results with radiative transfer simulations.

\section{Selection of the targets}
Our strategy was to find reliable GRB extinction curves in the literature and then select the host galaxies for which we were able to derive the attenuation curves. In this section, we briefly explain the methods to measure the extinction curve along GRB afterglow l.o.s in order and we define selection criteria to select the GRB extinction curves. The last step is the selection of the host galaxies.
\subsection{GRB selection}
\subsubsection{SED of GRB afterglows}
GRB afterglows are well described by synchrotron emission from accelerated electrons in a relativistic blast wave interacting with an external medium \citep{Sari98, Granot02}. The observed synchrotron spectrum is composed of series of power-law segments joined at certain break frequencies (see \citealt{Sari98, Granot02}). The NIR-to-X-ray spectrum is then usually modelled as a single power-law or a broken power-law to take into account a possible break in the synchrotron spectrum, most likely the cooling break, located between the X-ray and UV/optical/NIR ranges at the time of observation. 
Knowing the intrinsic spectral shape, the comparison of the observed spectrum with an expected one gives a direct access to the extinction curve. The non-attenuated intrinsic GRB afterglow SED is modelled by
\begin{ceqn}
\begin{equation}
    F_\nu = \begin{cases}
                  F_0 \cdot \nu^{-\beta_O}  & \nu \leq \nu_{break} \\
                  F_0 \cdot \nu_{break}^{\beta_X-\beta_O} \cdot \nu^{-\beta_X} & \nu > \nu_{break}
            \end{cases}
\end{equation}
\end{ceqn}
where $F_0$ is the flux normalisation, $\nu$ the frequency, $\nu_{break}$ the break frequency, $\beta_O$ and $\beta_X$ are the spectral slopes of the UV/optical/NIR and X-ray segments respectively. The free parameters are $F_0$, $\beta_O$ and $\beta_X$. At sufficiently late times the synchrotron emission is expected to radiate in the slow cooling regime, where $\beta_O = \beta_X + 0.5$ \citep{Sari98,Zafar11}.  
This emitted flux is then attenuated by dust absorption and scattering and photoelectric absorption of soft X-rays in the the host galaxy. The attenuated flux is given by
\begin{ceqn}
\begin{equation}
	F_{\nu}^{obs} = F_{\nu} \cdot 10^{-0.4 A_{\lambda}} \cdot e^{-N_{H,X} \sigma(\nu)}
\end{equation}
\end{ceqn}
where $A_{\lambda}$ is the wavelength dependent extinction in the host galaxy, $N_{H,X}$ is the equivalent neutral hydrogen column density derived from the soft X-ray absorption along the GRB l.o.s in the host galaxy and is a free parameter. $\sigma(\nu)$ is the cross-section for the photoelectric absorption from the gas in the host galaxy. 

\subsubsection{Modelling extinction processes}
\begin{figure}
	\centering
	\includegraphics[width=\hsize]{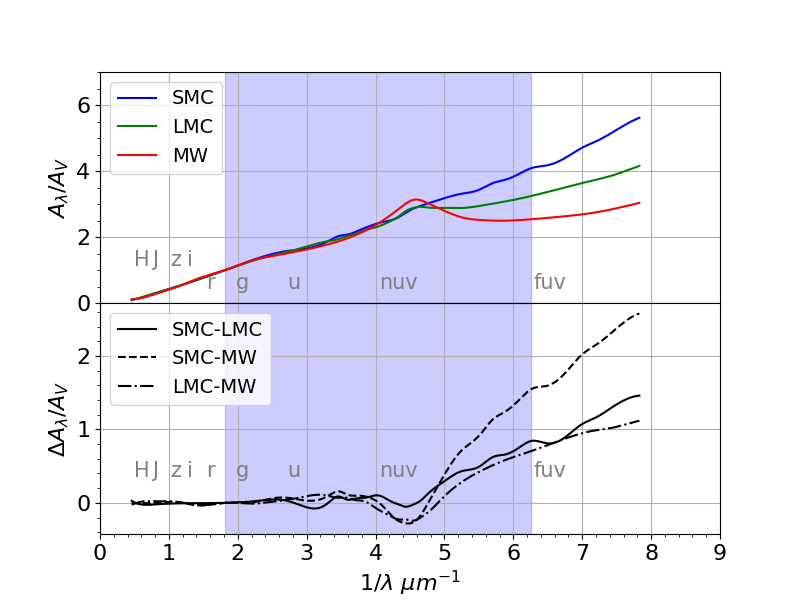}
	\caption{\textit{Top}: Mean local extinction curves using data of Table 1 in \cite{Pei92}, normalised to $A_V$. \textit{Bottom}: Difference between the mean local extinction curves, normalised to $A_V$. They start to differ significantly in the FUV. The blue shaded area represents the minimum rest-frame spectral coverage we require, that is, 1600\AA\ to 5500\AA\ .}
	\label{fig:Pei92}
\end{figure}

There are two methods to assess $A_{\lambda}$, in other words, the extinction curve. 
The first method consists in scaling the well defined extinction curves derived for the LMC, SMC, or MW to the data. The mean local extinction curves of \cite{Pei92}, normalised to the amount of extinction in the V band, $A_V$, are seen on top panel of Figure \ref{fig:Pei92}. These local laws are scaled to the GRB afterglow spectrum using $A_V$ as the scaling factor. This method has the advantage of introducing only one free parameter, $A_V$, to the SED fitting, but assumes that the wavelength dependence of dust extinction at high redshift is the same as the mean extinction in our local environment. Recently, Zafar et al. (2018, submitted) suggested avoiding use of the canonical \cite{Pei92} SMC curve to fit GRB data and to use the SMC-bar extinction curve from \cite{Gordon03}.\\
The second method is to perform a fit to assess directly the wavelength dependence of dust extinction. Usually the fit is performed using the prescriptions of \cite{Fitzpatrick07} (hereafter FM07) which parametrises the extinction curve as
\begin{ceqn}
\begin{equation}
	A_{\lambda} = A_V \left( \frac{1}{R_V} k(\lambda-V) + 1 \right)
\end{equation}
\end{ceqn}
where $\rm R_V = A_V/E(B-V)$ is the ratio of absolute-to-selective extinction.\\
In the UV wavelength range ($\lambda$ $<$ 2700\AA):
\begin{ceqn}
\begin{equation}
    k(\lambda-V) = 	\begin{cases}
                  	c_1 + c_2 x +c_3 D(x,x_0,\gamma) & x \leq c_5 \\
                  	c_1 + c_2 x +c_3 D(x,x_0,\gamma) + c_4 (x-c_5)^{2} & x > c_5
            			\end{cases}
\end{equation}
\end{ceqn}
where $x \equiv \lambda^{-1}$ in units of $\mu m^{-1}$, $c_1$ and $c_2$ represents the UV linear component, $c_3$ represents the bump strength, $c_4$ and $c_5$ the FUV curvature,  and the Drude profile modelling the 2175\AA\ bump is expressed as
\begin{ceqn}
\begin{equation}
D(x,x_0,\gamma) = {{ x^{2}} \over {(x^2-x_0^2)^2 + x^2 \gamma^2}}.
\end{equation}
\end{ceqn}
where $x_0$ is the central wavelength number and $\gamma$ is the width of the bump. In the optical and NIR a spline interpolation is used to estimate the extinction curve. The FM07 parametrisation offers more freedom to fit the actual shape of the extinction curve. However it has nine free parameters which leads to degeneracies between the parameters in the fitting process unless one has a finely-sampled SED. Consequently, this method is mainly used with spectroscopic measurements, as with photometric measurements only one has to fix some of the parameters, such as intrinsic spectral slope. Moreover, photometric observations affected by the IGM absorption are usually not used which further limits the number of data.\\

\subsubsection{Selection criteria for GRB extinction curves}

For our study purpose, it is essential to have a well characterised extinction curve. The three local extinction curves (MW, LMC, and SMC) have a very similar shape in the rest-frame NIR and optical and start to differ significantly in the rest-frame FUV (Figure \ref{fig:Pei92}). To be able to distinguish them, observations in the rest-frame UV bluewards of 2000\AA\ are needed. The lower panel of Figure \ref{fig:Pei92} shows the difference between the local extinction curves normalised to A$_V$. We can see that for low to moderate A$_V$ the difference in magnitudes between the local extinction curves may be below a standard photometric accuracy. A typical amount of extinction in the V band of 0.2 mag leads to a difference between the different local extinction curves higher than 0.15 mag only at 1600\AA\ in the rest-frame. It increases to at least 0.75 mag with $\rm A_V$=1 mag. In the following, the amount of extinction in the V band along the GRB l.o.s will be denoted $A_V^{grb}$. \\
We started by considering only GRB extinction curves derived using the FM07 formula with at least observations between 1600\AA\ and 5500\AA\ in the rest-frame. We used the largest sample of GRB extinction curves derived from spectroscopic measurements of the afterglow from \cite{Zafar11} and the updated results in \cite{Zafar12}. From the 42 GRBs of this sample, we kept the 31 GRBs with evidence of dust extinction along the GRB l.o.s, that is, $A_V^{grb}$ not consistent with 0 mag within 1$\sigma$ uncertainty. After applying the minimum wavelength coverage defined above we are left with a sample of eight GRBs: five with an SMC-like extinction curve and three exhibiting a 2175\AA\ bump (see Table \ref{tab:spec_sample}). Initially, GRB 080805 is not part of our sample due to the scarcity of optical data which can not safely constrain the extinction curve slope. However, the spectroscopic confirmation of the presence of a UV bump is an interesting information about the dust inside the galaxy and we decide to include GRB 080805 in our sample. To our knowledge there are only four GRBs with a spectroscopically confirmed UV bump (GRB 070802, GRB 080605, GRB 080607, GRB 080805) and they are all in our sample. 
\\
In order to increase our sample we also considered GRB extinction curves derived using the scaling method. We started with the sample of \cite{Japelj15} based on spectroscopic measurement of the afterglow. It consists of nine GRBs observed within the X-Shooter GRB GTO programme between 2009 and 2014. For the scaling method, the slope of the extinction curve is fixed, therefore we apply a relaxed minimum wavelength coverage going from 1600 to 3500 \AA\ in the rest-frame. We kept the eight GRBs with evidence of dust and after applying our wavelength coverage criterion we are left with four GRBs. GRB 120119A is best fitted with an LMC extinction curve, however \cite{Japelj15} reported that it overpredicts the strength of the UV bump and that none of the mean LMC, SMC and MW is able to accurately match the the binned spectrum of GRB 120119A. Using the afterglow spectrum observed by the X-Shooter and with the same data analysis as in \cite{Zafar11}, Zafar et al. (2018, submitted) find an SMC-like extinction curve using the FM07 formula. In the following we therefore consider the extinction curve to be SMC-like. We are left with 4 GRBs with an SMC-like extinction law. 
We also considered the sample of \cite{Schady12} based on broad-band photometry measurement of the afterglow since they required at least one detection at a rest-frame wavelength below 2000\AA. It consists of 49 GRBs. We kept the 36 GRBs with evidence of dust extinction and after applying our relaxed wavelength criterion we were left with 17 GRB extinction curves: eight consistent with an SMC extinction curve, six with an LMC one, and three with an MW one (see Table \ref{tab:spec_sample}).\\
At the end we have a sample of 30 GRB extinction curves: 17 with an SMC-like curve, six with an LMC-like, three with an MW-like, and four with a spectroscopically confirmed bump exhibiting various slopes.

\subsection{GRB Host galaxies selection}
Now that we have selected the extinction curves of 30 GRBs, we first compile some properties of their host galaxies, such as the SFR, stellar mass, and amount of attenuation in order to search for a correlation between the dust properties along the GRB l.o.s and inside the host galaxy. For the second part of our study we selected only the host galaxies for which we can derive the attenuation curves from SED fitting.

\subsubsection{GRB host galaxy properties}

The SFR derived from the dust corrected $\rm H_{\alpha}$ luminosity \citep{Kennicutt98} is considered as one of the most robust tracers of recent star formation, in the following it will be denoted $\rm SFR_{ H_{\alpha}}$. 
The extinction of emission lines derived from the Balmer decrement is an indicator of the dust extinction towards the HII regions. In the following, the amount of attenuation in the V band in the host derived from the Balmer decrement will be denoted $\rm A_{V}^{lines}$. For homogeneity, these two quantities are exclusively taken from \cite{Kruehler15}, with the exception of GRBH 061121 for which the $\rm SFR_{H_{\alpha}}$ is taken from \cite{Vergani17} and $\rm A_{V}^{lines}$ is computed from the H$_{\alpha}$ and H$_{\beta}$ fluxes in \cite{Vergani17} using the same prescriptions as in \cite{Kruehler15}. The metallicities were measured using the calibration of \cite{Maiolino08}, and taken from \cite{Kruehler15}, \cite{Japelj16} and \cite{Vergani17}. The stellar masses are taken from \cite{Perley16}.

\subsubsection{Golden sample for deriving GRBH attenuation curves}
The characterisation of the dust effects is crucial in the UV. From the 30 GRB host galaxies selected with robust GRB extinction curves we select only targets with photometric measurements of the UV rest-frame emission of the host and a UV-to-NIR spectrum reasonably well sampled (with at least six bands) in order to measure physical parameters.
After a careful inspection of the photometric and spectroscopic data of the host galaxies of GRBs selected in section 2.2.1, we select seven galaxies with good photometric data on at least six bands, including a detection with the Spitzer Infrared Array Camera at 3.6$\mu m$ (IRAC1) and hydrogen recombination lines (except for GRBH 080607 at z=3.04) allowing a measure of $\rm SFR_{ H_{\alpha}}$ and of the Balmer decrement. The photometric data collected for each target are described in Appendix B. The data are corrected for the foreground Galactic extinction using \cite{schlafly11}. The seven hosts are gathered in Table \ref{targets}.

\begin{table}
	\renewcommand{\arraystretch}{1.5}
	\small
	\caption{ GRB hosts in our Golden sample. The only host which does not meet our selection criteria of minimum wavelength coverage is listed at the bottom.}
	\label{targets}
	\centering
	\begin{tabular}{cccc}
	\hline
	Name & redshift & Ext. curve & Host data \\   
	\hline
	\hline
	061121$^{(a)}$ & 1.31 & LMC & 10 bands (incl. VLA)\\
    070802 & 2.45 & relatively steep & 7 bands \\
           &      & + UV bump    & \\
	080605 & 1.64 & steep + UV bump & 8 bands \\
	080607 & 3.04 & flat + UV bump & 9 bands (incl. ALMA)\\
	100814 & 1.44 & SMC-like & 6 bands \\
	120119 & 1.72 & SMC-like & 6 bands\\
	\hline
	080805 & 1.51 & flat + UV bump & 6 bands \\
	\hline
	\end{tabular}
	\tablefoot{$(a)$: GRB 061121 is the only afterglow for which the extinction curve is derived with photometric data only.}
\end{table}

\section{GRB and Host properties}

Before comparing the extinction curve derived from the GRB l.o.s with the attenuation curve derived for the host galaxy we are searching for some correlations between properties derived along the GRB l.o.s and for the whole galaxy to check whether GRB l.o.s are representative for the dust properties of the hosts.

\subsection{Importance of a rapid follow-up}

\begin{figure}
	\centering
	\includegraphics[width=\hsize]{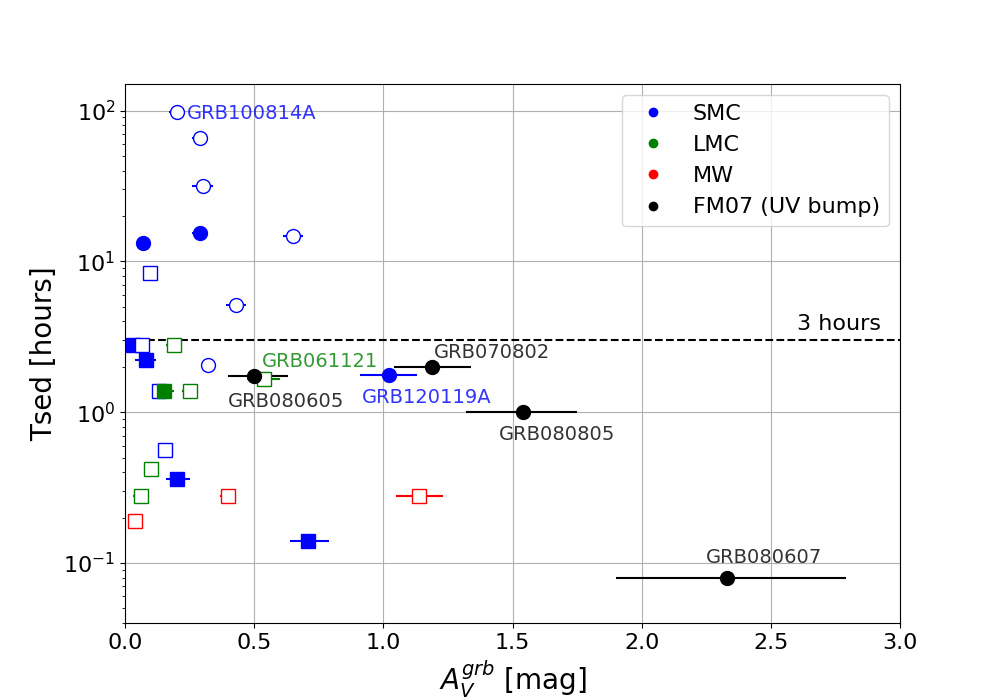}
	\caption[Caption for LOF]{Time, since the GRB trigger, at which the SED were evaluated as a function of the extinction along the GRB l.o.s, $\rm A_{V}^{grb}$. The blue, green and red colours represent which local extinction curve (SMC, LMC, MW) scaled the best to the data. The black colour means that the extinction curve was fitted with the FM07 and has a spectroscopically confirmed UV bump. The circles represent the afterglows with spectroscopic measurement and the squares the ones with photometric measurement only. The filled symbols represent SEDs best fitted with a single power law, the empty ones represent SEDs best fitted by a broken power law. Only the names of the GRB in our Golden sample are indicated}
	\label{fig:Tsed_Avgrb}
\end{figure}

As GRB afterglows fade rather rapidly, we first check whether the time at which the GRB SED is extracted has an influence on the study of GRB extinction curves. Figure \ref{fig:Tsed_Avgrb} represents the time, after the GRB trigger, at which the GRB SED is extracted as a function of $A_V^{grb}$ for different types of extinction curves. We see that at early times, below roughly three hours, GRB afterglows exhibit all type of extinction curves from low to high extinction whereas after three hours only SMC-like extinction curves with low to moderate extinction are found. There are two ways to explain this trend, either as a bias against highly extinguished afterglows or a time dependent dust composition modification. A temporal evolution of the dust properties is expected from theoretical simulations 
\citep{Waxman00,Fruchter01,Perna03} where a strong X-ray/UV radiation field can deeply modify the dust properties in the close environment ($\sim 10-100$ pc) of the source. Dust sublimation is predicted to affect primarily the smaller grains, and the silicates are preferentially destroyed compared to graphite, leading to featureless and shallow extinction curves. A possible early colour variation has been studied in \cite{Morgan14}. However, these processes mainly take place during the first seconds or minutes after the burst and not a few hours after the burst.\\
The other explanation is a bias against highly extinguished afterglows due to instrument sensitivity and fast decay of the GRB afterglow brightness. Indeed, if we consider that GRBs in dusty galaxies are not intrinsically brighter than those in galaxies with low dust content, they will be below instrument sensitivities very rapidly, with the exception of extremely bright afterglows. As mentioned in \cite{Zafar11}, where the afterglows in their sample have an average magnitude of 21.1$\pm$1.7, a MW-like extinction curve with a rest-frame visual extinction of $\rm A_{V}^{grb}$=1 mag will yield roughly 2 to 3.5 magnitudes of extinction in the optical at z=2. Therefore GRB afterglows with $\rm A_{V}^{grb}$>1 mag are unlikely to be detected with current instrument sensitivity except for the few brightest ones and those with a follow-up within the first minutes after the GRB trigger. The fraction of dark bursts has been estimated between 25 to 42\% in \cite{Fynbo09} and using the complete BAT6 sample, \cite{Melandri12} derived a 25-35\% fraction of genuine optically dark bursts. Using systematic follow-up observations from GROND, \cite{Greiner11} show that at moderate redshift, the main contribution to dark bursts is a moderate extinction. We conclude that the trend observed in Figure \ref{fig:Tsed_Avgrb} is very likely driven by the observed GRB brightness and not due to a temporal evolution of dust properties.\\
In Figure \ref{fig:Tsed_Avgrb}, we also see that extinction curves exhibiting a 2175\AA\ bump are preferentially found in the most extinguished l.o.s as previously found in \citep{Zafar11,Greiner11,Kruehler11}. They are therefore more easily detected at early times. It follows that there is an observational bias towards SMC-like extinction curves.\\
The forthcoming French Ground Follow-up Telescope (F-GFT), a 1.3m optical/NIR robotic telescope for the Space Variable Object Monitor mission (\citealt{SVOM16}), with its extremely fast response time of less than 20s after receiving an alert, should significantly increase the detection of extinguished bursts. It should also be able to observe at early times the rapid afterglow colour change characteristic of a temporal variation of dust properties under the influence of intense X-ray/UV radiation field.  

\footnotetext{For GRBs from \cite{Zafar11,Zafar12,Schady12} this is the time at which the SED is extracted, and for those from \cite{Japelj15} the time corresponds to the mid-time of the X-shooter observations.}

\begin{figure}
	\centering
	\includegraphics[width=\hsize]{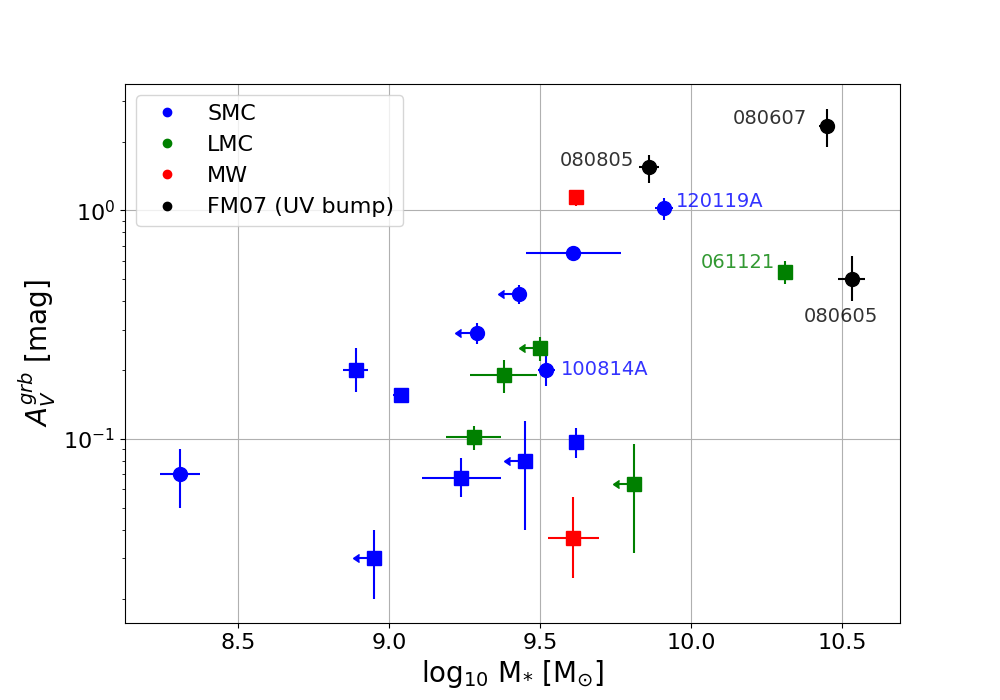}
	\caption{Extinction along the GRB l.o.s as a function of the host galaxy stellar mass, the latter taken from \cite{Perley16}. Same colours and symbols as in Figure \ref{fig:Tsed_Avgrb}}
	\label{fig:Avgrb_M}
\end{figure} 
\subsection{Dust extinction along GRB l.o.s as a function of GRBH stellar mass}

Figure \ref{fig:Avgrb_M} represents $A_V^{grb}$ as a function of the GRBH stellar mass for different types of extinction curves. It shows that highly extinguished GRB l.o.s are preferentially found in the most massive host galaxies. Previous studies (e.g. \citealt{Zahid13, Heinis14}) showed that the amount of dust attenuation in a galaxy is correlated with the stellar mass of the galaxy. \cite{Vijh03} found a similar relation between the UV attenuation and UV luminosity of Lyman break galaxies. This relation holds if the GRB afterglow probes not only the dust capable to survive the intense radiation of the afterglow in the surrounding environment of the GRB progenitor but also the dust located at sufficiently large distances to be representative of the whole galaxy. Interestingly, this figure also shows that GRB extinction curves exhibiting a UV bump are preferentially found in the most massive host galaxies as already noticed in \cite{Kruehler11}. \cite{Noll09b} find that galaxies with evidence of a UV bump have a more evolved stellar population than galaxies lacking this feature. 
\section{Fitting the SED of the hosts}

For the seven GRBHs in our Golden sample defined in Table \ref{targets} we perform SED fitting to derive the attenuation curve. The method and the models are first described, then we discuss the SED fitting analysis of each GRB host galaxy. In the last part of this section we discuss the star formation activity of these seven GRBHs.
\begin{table*}
	\renewcommand{\arraystretch}{1.5}
	\caption{Input parameters for SED fitting with CIGALE }
	\centering
	\begin{tabular}{c c c}
	\hline
	Parameter & Symbol & Range \\
	\hline\hline
	Colour excess \tablefootmark{1} & $\rm E(B-V)_{s}$ & 0.01,0.05, 0.1,0.25,0.5,0.75,1.0,... \\
	& & 1.25,1.5,1.75,2.0,2.25,2.5,3.0,4.0 mag \\
	Attenuation curve\tablefootmark{2}& $\delta$ & [-1:+0.7] per bin of 0.1\\
	\hline
	{Delayed star formation history $+$ recent burst }\\
	\hline
	Age of the main stellar population & $t_0$ &  1000, 2000, 3500\,Myr \\
	Peak of the star formation (delayed SFH) & $\tau$ &  1000, 2000, 3000, 5000, 7000, 10000  Myr\\
	Age of the burst & $t_{\rm b}$ &  20, 50, 100 \,Myr \\
	Stellar mass fraction due to the burst & $f_{\rm b}$ &  [0:0.5] per bin of 0.05\\
	Metallicity\tablefootmark{3} & $Z$ &  0.008, 0.02\\
	\end{tabular}
	\tablefoot{Values of input parameters used for the SED fitting with CIGALE. \\
	\tablefootmark{1}{$\rm E(B-V)_{s}$ is the colour excess between the B and V bands applied on the stellar continuum.}\\
	\tablefootmark{2} { Defined in Eq.\ref{eq:attlaw}}\\
	\tablefootmark{3} {The metallicity is set to one of the 2 values ($Z$=0.02 being the solar metallicity)}\\
	}
	
	\label{tab:parameter}
\end{table*}

\subsection{SED fitting method}
\label{sec:SED_fitting}
The SED fitting is performed with the  version v0.11 of the CIGALE code (Code Investigating GALaxy Emission)\footnote{http://cigale.lam.fr}. CIGALE combines a UV-optical stellar SED with a dust component emitting in the IR and fully conserves the energy balance between  dust absorbed  emission and its re-emission in the IR. Star formation histories as well as dust attenuation characteristics, including the attenuation law, are input parameters that can be either taken free or fixed according to the available data or the specific aims. The main characteristics of the code are described in \cite{noll09}. We refer to  Boquien et al. 2017 (in preparation) for a detailed description of the new version of the code and to the online information to use the code. Here we only describe the assumptions and choices specific to the current study. The main parameters and range of input values are reported in Table \ref{tab:parameter}. The output values of the parameters are estimated by building the probability distribution function (PDF) and by taking its mean and standard deviation. To assess the quality of the fit we compare the results to the SFR derived from the dust corrected $\rm H_{\alpha}$ luminosity and the amount of dust attenuation derived from the Balmer decrement.\\
Different scenarios of star formation history are implemented in CIGALE. In this work, we have adopted a delayed star formation rate ($\rm SFR \propto t \cdot e^{-t/\tau}$) on which we add a young burst of constant star formation with a free age and amplitude. This scenario aims at reproducing both the general star formation history of distant galaxies (i.e. \citealt{Lee10, Maraston10, Pforr12}) and a possible recent burst of star formation. The hosts of GRBs are known to be active in star formation (\citealt{Savaglio09,Perley13,Hunt14,Japelj16}). As rest-frame UV photometric data can probe down to the last 20 Myr of star formation history (\citealt{Boquien14}), the age of the burst is allowed to vary from 20 Myr to 100 Myr. The fraction of stellar mass in the recent burst is allowed to vary from 0 to 50\% of the main stellar population. \\
The initial mass function of \cite{chabrier03} is adopted with the stellar synthesis models of  \cite{bc03}. The nebular emission is included with standard parameters. The metallicity is fixed and chosen to be the closest value to the measurements between Z=0.008 and 0.02 (solar value).\\
To model the attenuation by dust, we used a modified version of the attenuation law of C00, which offers the possibility of varying the steepness of this law and adding a bump centred at 2175\AA . We refer to \cite{noll09} for a complete description of the dust attenuation prescription, which still holds in the new version of the code. In brief, the dust attenuation is described as
\begin{ceqn}
\begin{equation}\label{eq:attlaw}
A_\lambda  = {E(B-V)_s}~ \left[ k'(\lambda) \left(\lambda \over {\lambda_V} \right)^{\delta} + D_{\lambda_0,\gamma,E_ {\rm b}}(\lambda) \right]
\end{equation}
\end{ceqn}
where $\lambda_V = 5500$\AA\ , $k'(\lambda)$ comes from C00. $D_{\lambda_0,\gamma,E_{\rm b}}(\lambda)$, the Lorentzian-like Drude profile commonly used to describe the UV bump \citep{Fitzpatrick90}, is defined as
\begin{ceqn}
\begin{equation}\label{eq:bump}
D_{\lambda_0,\gamma,E_{\rm b}} = {{E_{\rm b} \lambda^2 \gamma^2} \over {(\lambda^2-\lambda_0^2)^2 + \lambda^2 \gamma^2}}.
\end{equation}
\end{ceqn}
with $\lambda_0 =$ 2175\AA\ .
An extinction curve as steep as the SMC corresponds to $\delta\simeq -0.5$ and the bump amplitude found in the average extinction curve of the MW corresponds to $\rm E_{b}$=3.5. For a sample of UV selected galaxies at intermediate redshift \cite{Buat12} find $\rm E_{b} = 1.6$ and $\delta=-0.27$, which is close to the  shape of the LMC2 extinction curve with a UV bump of moderate amplitude.\\
The photometric data partially overlap the bump for GRBHs 061121, 070802, 080605, 080805, 100814A and 120119A. For these six sources the  bump will be fixed at $\lambda_0= 2175$\AA\ with a width $\gamma= 350 $\AA\ and the amplitude is taken free. Because of the lack of wavelength coverage the other fits are performed without adding any bump in the attenuation curve and only varying its slope ($\delta$).\\
The stellar emission absorbed by dust is re-emitted in the IR. The IR templates used in this work  are from \cite{dale14} (without the AGN component) which are parametrised with a single parameter $\alpha$ corresponding to the exponent of the distribution of heating intensity over dust mass. A single template ($\alpha$ = 2) is used except for the only galaxy detected in IR, GRBH 080607, for which different $\alpha$ of between 1 and 2 are tested to cover the range of values found from quiescent to active star-forming galaxies (\citealt{dale02}).\\
The attenuation of the continuum obtained from the SED fitting will be denoted $\rm A_{V}^{stars}$.

\subsection{Analysis of the sources}
 
\subsubsection{GRBH 061121}

\begin{figure}
	\centering
	\includegraphics[width=\columnwidth]{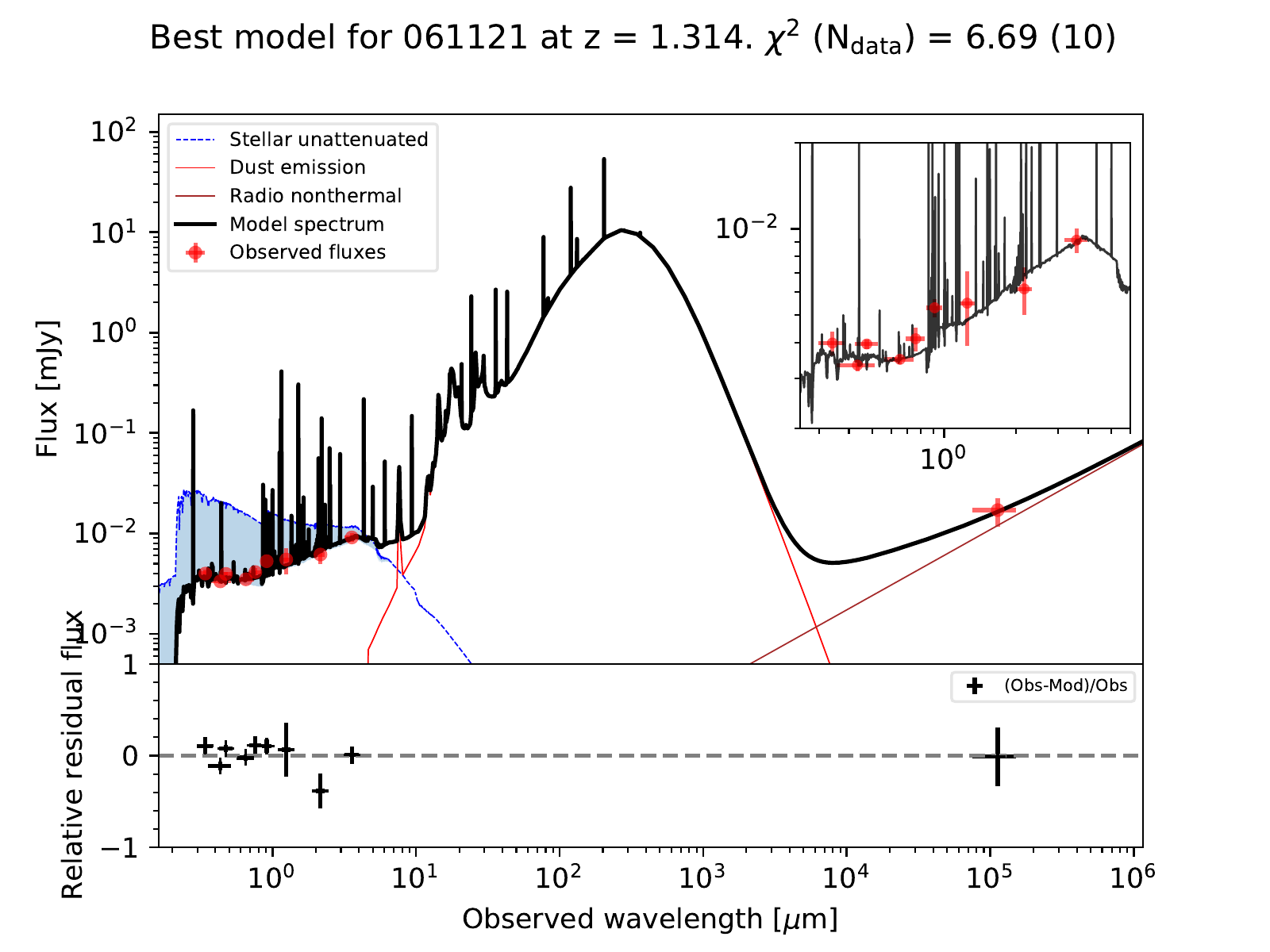}
	\caption{Best-fitted SED for GRBH 061121. Red points are the observed fluxes with their uncertainties. The black line represents the best fitted spectrum composed of attenuated stellar emission, dust emission, non-thermal radio emission and nebular lines. The dashed blue line represents the unattenuated stellar emission and the blue shaded area indicates the amount dust attenuation. The lower panel shows the relative residual. The upper right panel is an enlargement of the UV/optical/NIR domain.}
	\label{best061121}%
\end{figure}

From $\rm H_{\alpha}$ measurements, \cite{Vergani17} derived a SFR$_{\rm H\alpha}$ of 44.2$^{+19}_{-10}$ $\rm M_\sun yr^{-1}$. Using the Balmer lines measurements by \cite{Vergani17}, we computed $\rm E(B-V)_g = 0.53^{+0.19}_{-0.19}$ mag corresponding to $\rm A_V^{lines} = 1.66^{+0.60}_{-0.60}$ mag (using $R_V$=3.08 as in \citealt{Kruehler15}). From IRAC1 measurement \cite{Perley16} derived a stellar mass of $\rm 2 ~10^{10} M_\sun$.\\
The radio component is added to the fit with standard parameters (a FIR/radio correlation coefficient of 2.58 and a slope of 0.8 for the power-law synchrotron emission). 
The best model is obtained for Z=0.008 and shown in Figure \ref{best061121}. No evidence of a UV bump is found.
The SFR is found equal to $\rm 62 \pm 26~ M_\sun yr^{-1}$ consistent with SFR$_{\rm H\alpha}$. A strong burst involving 29$\pm 14 \%$ of the total mass with an intermediate age of 58$\pm 31$ Myrs is needed to reproduce the SED. The stellar mass is $\rm M_\star = 6.9 \pm 2.4 ~10^ 9 M_\sun$ a factor 3 lower than the one measured by \cite{Perley16}. The resulting specific SFR, defined as $\rm sSFR=SFR/M_{*}$ is equal to $\rm 9 ~Gyr^{-1}$. The dust attenuation is found  moderate with $\rm E(B-V)_s=0.15\pm 0.04$ mag, with a flat attenuation law ($\delta= 0.22\pm 0.2$) marginally consistent with the starburst law of C00. The resulting amount of visual extinction is $\rm A_{V}^{stars} = 0.88 \pm 0.30$ mag. 
A metallicity of Z=0.02 returned a slightly higher $\chi^2$ with similar values for the SFR, the stellar mass, $\delta$ and $\rm A_{V}^{stars}$.\\
\subsubsection{GRBH 070802}
\label{sec:070802}
\begin{figure}
	\centering
	\includegraphics[width=\columnwidth]{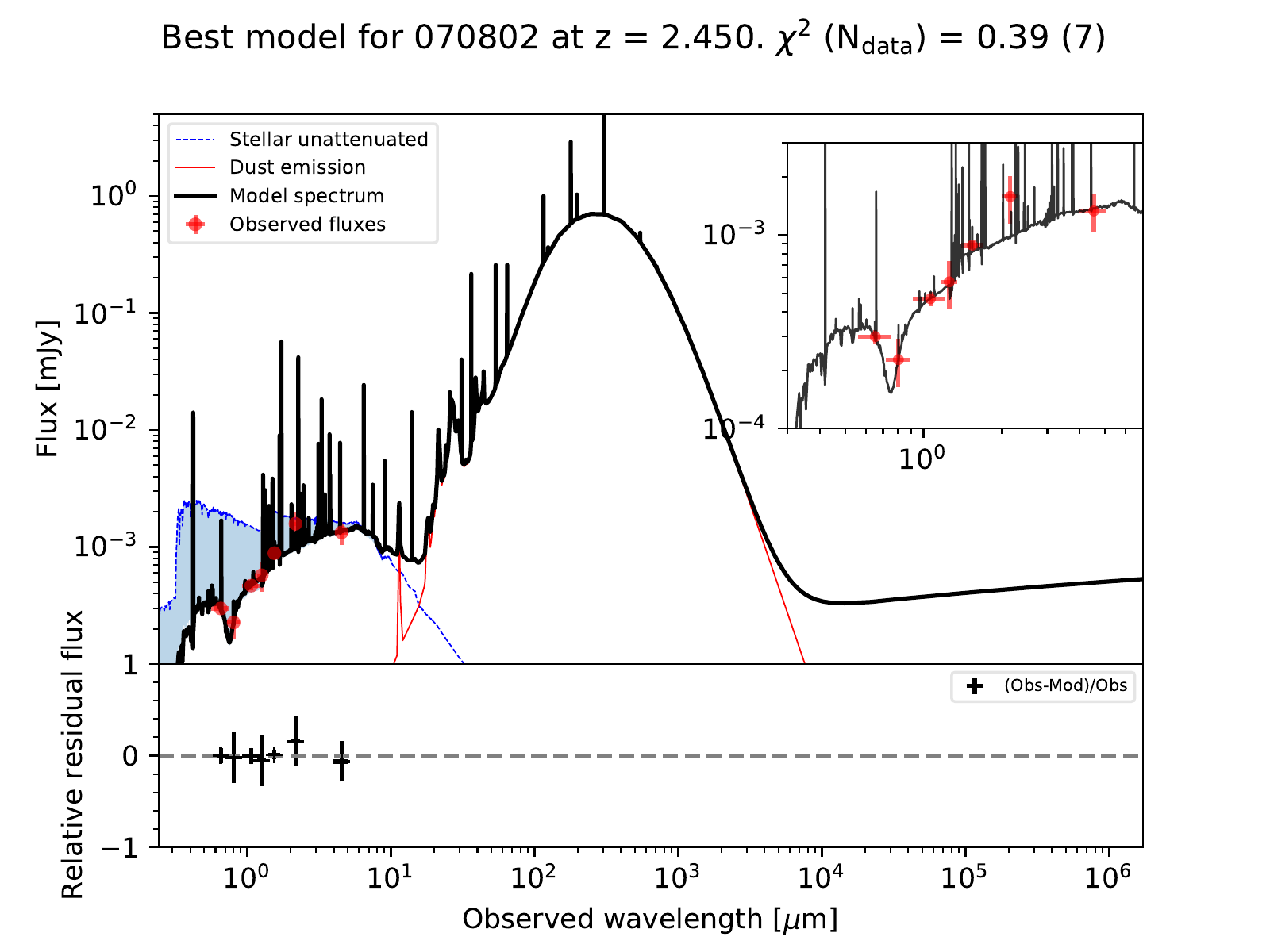}
	\caption{Best-fitted SED for GRBH 070802. Colour and symbols convention as in Figure \ref{best061121}.}
	\label{best070802}%
\end{figure}

The Balmer lines are measured by \cite{Kruehler15}, corresponding to a moderate attenuation with  $\rm E(B-V)_g = 0.31^{+0.12}_{-0.12}$ mag corresponding to $\rm A_V^{lines}$ = $0.95^{+0.37}_{-0.37}$ mag and $\rm SFR_{H\alpha} = 24^{+11}_{-8}~ M_\sun yr^{-1}$.\\
The best model is obtained for Z=0.008 and shown in Figure \ref{best070802}.
The SFR is found equal to $\rm 32.2 \pm 17.8~ M_\sun yr^{-1}$ consistent with SFR$_{\rm H\alpha}$. A strong burst involving 26$\pm 15 \%$ of the total mass with an intermediate age of 55$\pm 32$ Myrs is needed to reproduce the SED. The stellar mass is $\rm M_\star = 3.7 \pm 1.6 ~10^ 9 M_\sun$. \\
The dust attenuation is found  moderate with $\rm E(B-V)_s=0.21\pm 0.05$ mag, with a relatively steep attenuation law ($\delta= -0.12\pm 0.30$) consistent within 1$\sigma$ uncertainty with the starburst law of C00. The resulting amount of visual extinction is $\rm A_{V}^{stars} = 0.86 \pm 0.35$ mag. Letting the UV bump amplitude vary between 0 and 6, as observed in MW sighlines, a significant bump is found with $\rm E_b = 3.6 \pm 1.7$. This detection is due to the flux drop observed in the \textit{i'} band, the large uncertainty does not allow to tightly constrain the bump amplitude ($i'=25.5 \pm 0.3$ mag). As the PDF of the bump amplitude is cut close to the peak of the PDF, we have performed the fit again by letting the bump amplitude to vary between 0 and 10 and it results in $\rm E_b = 5.3 \pm 2.6$ with a better defined PDF, while the $\chi^2$ and estimations for the other output parameters remain unchanged. We also performed the fit without bump and the estimation of the SFR, $\rm M_\star$ and $\rm E(B-V)_s$ remain unchanged while the slope of the attenuation curve becomes steeper with $\delta=-0.31 \pm 0.26$. with a higher $\chi^2$.
A metallicity of Z=0.02 returned a slightly higher $\chi^2$ with similar values for the SFR, the stellar mass, $\delta$ and $\rm A_{V}^{stars}$.\\
\subsubsection{GRBH 080605} 

The Balmer lines are measured by \cite{Kruehler15}, corresponding to a moderate attenuation with  $\rm E(B-V)_g = 0.26^{+0.11}_{-0.10}$ mag and $\rm A_V^{lines} = 0.8^{+0.31}_{-0.34}$ mag with an intense star formation: SFR$_{\rm H\alpha}$ = 47$^{+17}_{-12}$ $\rm M_\sun yr^{-1}$. From IRAC1 measurement, \cite{Perley16} derive a stellar mass of $\rm 3.4 ~10^{10} ~M_\sun$. \\
\begin{figure}
	\centering
	\includegraphics[width=\columnwidth]{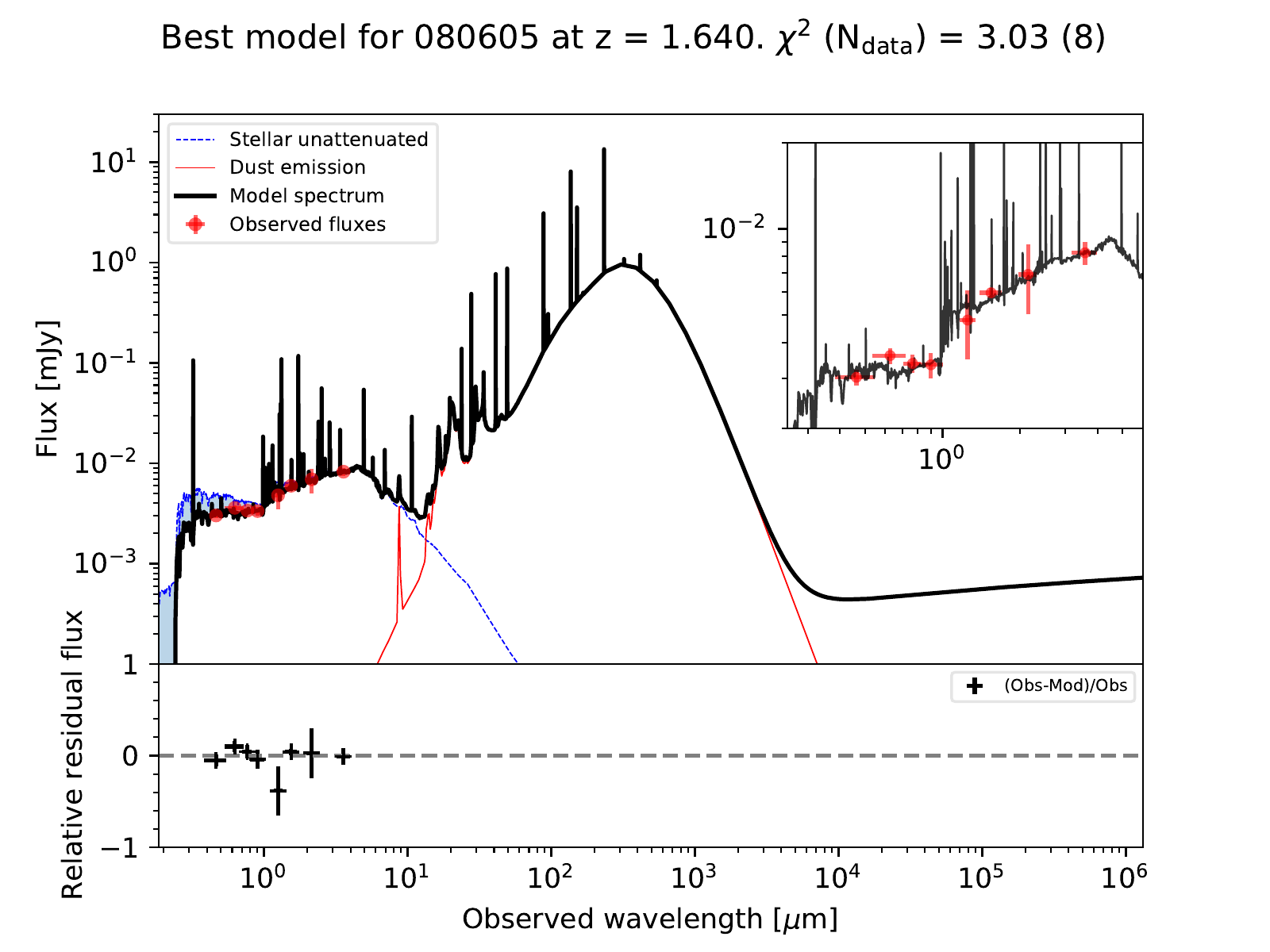}
	\caption{Best-fitted SED for GRBH 080605. Colour and symbols convention as in Figure \ref{best061121}.}
	\label{best080605}%
\end{figure}
The best model is obtained for Z=0.02 and shown in Figure \ref{best080605}. 
The SFR found equal to $\rm 44.9 \pm 22.9~ M_\sun yr^{-1}$ is in very good agreement with the one measured with the H$\alpha$ flux. A recent burst of star formation is found with an amplitude of 15$\pm 14 \%$ of the total mass and an age of 75$\pm 31$ Myrs. The stellar mass is $\rm M_\star = 12.3 \pm 4.2 ~10^ 9 ~M_\sun$  a factor  2.8  lower than the one measured by \cite{Perley16}. We estimate a sSFR for this starbursting galaxy of $\rm 3.6 ~Gyr^{-1}$.\\
Dust attenuation is characterised by a colour excess $\rm E(B-V)_s = 0.09 \pm 0.05$ mag and a shallow attenuation curve consistent with C00 ($\delta=-0.08 \pm 0.38 $). This results in a modest amount of visual extinction of $\rm A_{V}^{stars} = 0.41 \pm 0.29$ mag. No evidence for a UV bump is found.
A metallicity of Z=0.008 returned a higher $\chi^2$ with similar values for the SFR, the stellar mass, $\delta$ and $\rm A_{V}^{stars}$.

\subsubsection{GRBH 080607}

The stellar mass is measured with IRAC1 data by \cite{Perley16}: $\rm M_\star = 2.8 ~10^{10} ~M_\sun$. No measure of the Balmer lines is available. \\
\begin{figure}
	\centering
	\includegraphics[width=\columnwidth]{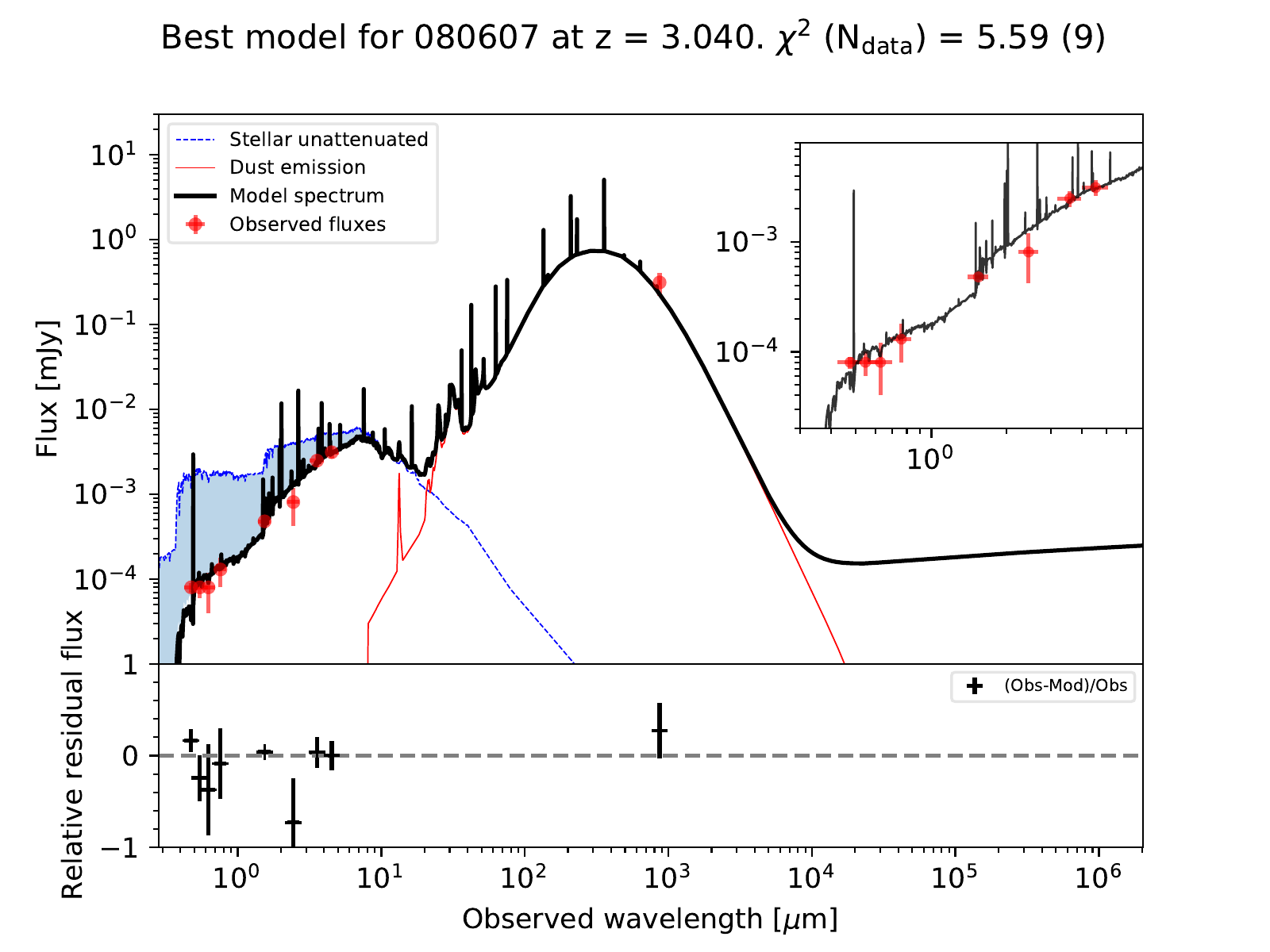}
	\caption{Best-fitted SED for GRBH 080607. Colour and symbols convention as in Figure \ref{best061121}.}
	\label{best080607}%
\end{figure}
We adopted a solar metallicity (Z=0.02) as measured for the host ISM from afterglow absorption-line observations \citep{Prochaska09}. With only one measurement for the dust emission, we cannot run CIGALE with a free dust component. Instead we performed several runs varying the $\alpha$ parameter ($\alpha=$ 1, 1.5 and 2), the quality of the fits are similar. Varying $\alpha$ changes the dust temperature and the position of the peak of the dust component. The IR luminosity, SFR, stellar mass and attenuation vary accordingly: $\rm L_{IR}$, SFR, $\rm M_{*}$ and $\rm A_V^{stars}$ decrease from $\rm \sim 7.2~10^{11} ~L_\sun$, $\rm \sim 70 ~M_\sun yr^{-1}$, $\rm \sim 3.4 ~10^{10} ~M_\sun$ and $\sim 2$ mag to $\rm \sim 1.9~10^{11} ~L_\sun$, $\rm \sim 20 ~M_\sun yr^{-1}$, $\rm \sim 2.1 ~10^{10} ~M_\sun$ and $\sim 1$ mag when $\alpha$ increases from 1 to 2. As expected the stellar mass remains similar. Taking in mind this uncertainty we fix $\alpha=1.5$, describing a galaxy extremely active in star formation.
The best fit is shown in Figure \ref{best080607}.\\
The stellar mass fraction due to the recent burst is 8$\pm$11\%, consistent with no burst, consequently we decide to report the results without any recent burst. The total IR luminosity of the galaxy is $\rm L_{IR} = 3.6 \pm 1.4~ 10^{11} ~ L_\sun$\ with a SFR equal to $\rm 35.2 \pm 13.9~ M_\sun yr^{-1}$. The stellar mass, $\rm M_\star = 2.75 \pm 0.8 ~10^{10} ~M_\sun$ is in very good agreement with the one measured by \cite{Perley16}. The sSFR is equal to $\rm 1.3 ~Gyr^{-1}$.
The dust attenuation is characterised by a colour excess of $\rm E(B-V)_s = 0.31 \pm 0.04$ mag and a flat attenuation curve ($\delta=0.05 \pm 0.10 $) consistent with C00. The resulting amount of visual extinction is $\rm A_{V}^{stars} = 1.39 \pm 0.29$ mag.\\
The fit with a recent burst gives a slightly higher SFR of $\rm 49.7 \pm 14~ M_\sun yr^{-1}$, a slightly lower stellar mass, $\rm M_\star = 2.3 \pm 1 ~10^{10}~M_\sun$, and similar colour excess and slope for the dust attenuation.

\subsubsection{GRBH 080805}

The Balmer lines are measured by \cite{Kruehler15}. They derive a high  colour excess although with a large uncertainty, $\rm E(B-V)_g = 0.78^{+0.39}_{-0.31}$ mag corresponding to an attenuation $\rm A_V^{lines}$ of  $2.4^{+1.2}_{-0.9} $ mag, and a highly uncertain star formation rate, $\rm SFR_{H\alpha} = 45^{+79}_{-26} M_\sun yr^{-1}$. The stellar mass derived from IRAC1 measurement by \cite{Perley16} is $\rm M_\star = 7.2  ~10^ 9 ~M_\sun$.\\
\begin{figure}
	\centering
	\includegraphics[width=\columnwidth]{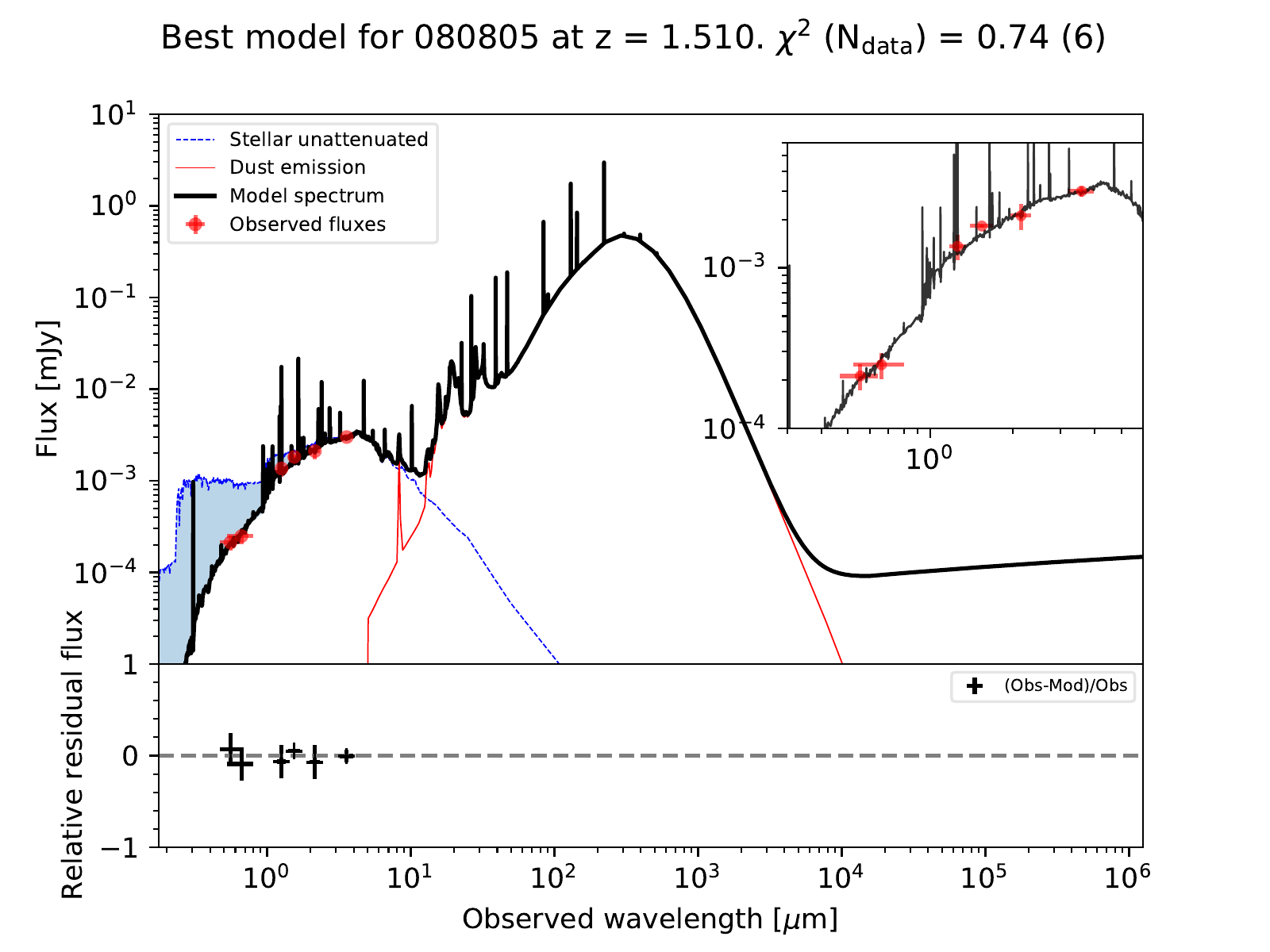}
	\caption{Best-fitted SED for GRBH 080805. Colour and symbols convention as in Figure \ref{best061121}.}
	\label{best080805}
\end{figure} 
The best model is obtained for Z=0.02 and shown in Figure \ref{best080805}. 
The SFR is found equal to $\rm 20.6 \pm 12~ M_\sun yr^{-1}$ is in agreement with the one measured with the H$\alpha$ flux within errorbars. A burst of star formation is found with an amplitude of $22 \pm 16 \%$ of the total mass and an age of $60 \pm 33$ Myrs.
The stellar mass is $\rm M_\star = 3.4 \pm 1.7 ~10^ 9 ~M_\sun$. We estimate a sSFR for this galaxy of $\rm 6 ~Gyr^{-1}$. The dust attenuation is characterised by a colour excess $\rm E(B-V)_s = 0.37 \pm 0.07$ mag and a steep attenuation curve, similar to that of the SMC, ($\delta=-0.45 \pm 0.24 $). This results in an amount of visual extinction of $\rm A_{V}^{stars} = 1.06 \pm 0.34$ mag. No evidence for a UV bump is found.
A metallicity of Z=0.008 returns a worse $\chi^2$ with similar values for the SFR and stellar mass but different values for the dust attenuation: $\delta$ ($-0.19 \pm 0.23 $) and $A_{V}^{stars}$ ($1.38 \pm 0.47$ mag). \\
\\
In order to improve the agreement between the SFR found from the SED and from H$\alpha$ measurements, we checked whether the presence of a very young starburst could enhance the current star formation rate. We ran again CIGALE with a recent burst taking place within the last 10 to 20 Myrs. The SFR increases to $\rm 48.2 \pm 29.9 ~ M_\sun yr^{-1}$, in very good agreement with the H$\alpha$ measurements. The burst of star formation is found with an amplitude of $22 \pm 16 \%$ of the total mass and an age of 15$\pm 4$ Myrs. As expected, the stellar mass is reduced to $\rm M_\star = 2.9 \pm 1.6 ~10^9 ~M_\sun$. The attenuation curve becomes shallower, but still steeper than the one of C00, with $\delta=-0.25 \pm 0.24$ and the amount of attenuation in the optical increases to $\rm A_V^{stars}= 1.53 \pm 0.5 $ mag. This indirect use of the H$_{\alpha}$ emission line indicates that a very recent episode of star formation can explain the  measured SFR$\rm _{H\alpha}$. In a future work, taking into account the emission lines in the fit would allow to probe star formation at timescales lower than 20Myrs. We retain the result with a burst age older than 20Myrs as the reference.

\subsubsection{GRBH 100814A}
\begin{figure}
	\centering
	\includegraphics[width=\columnwidth]{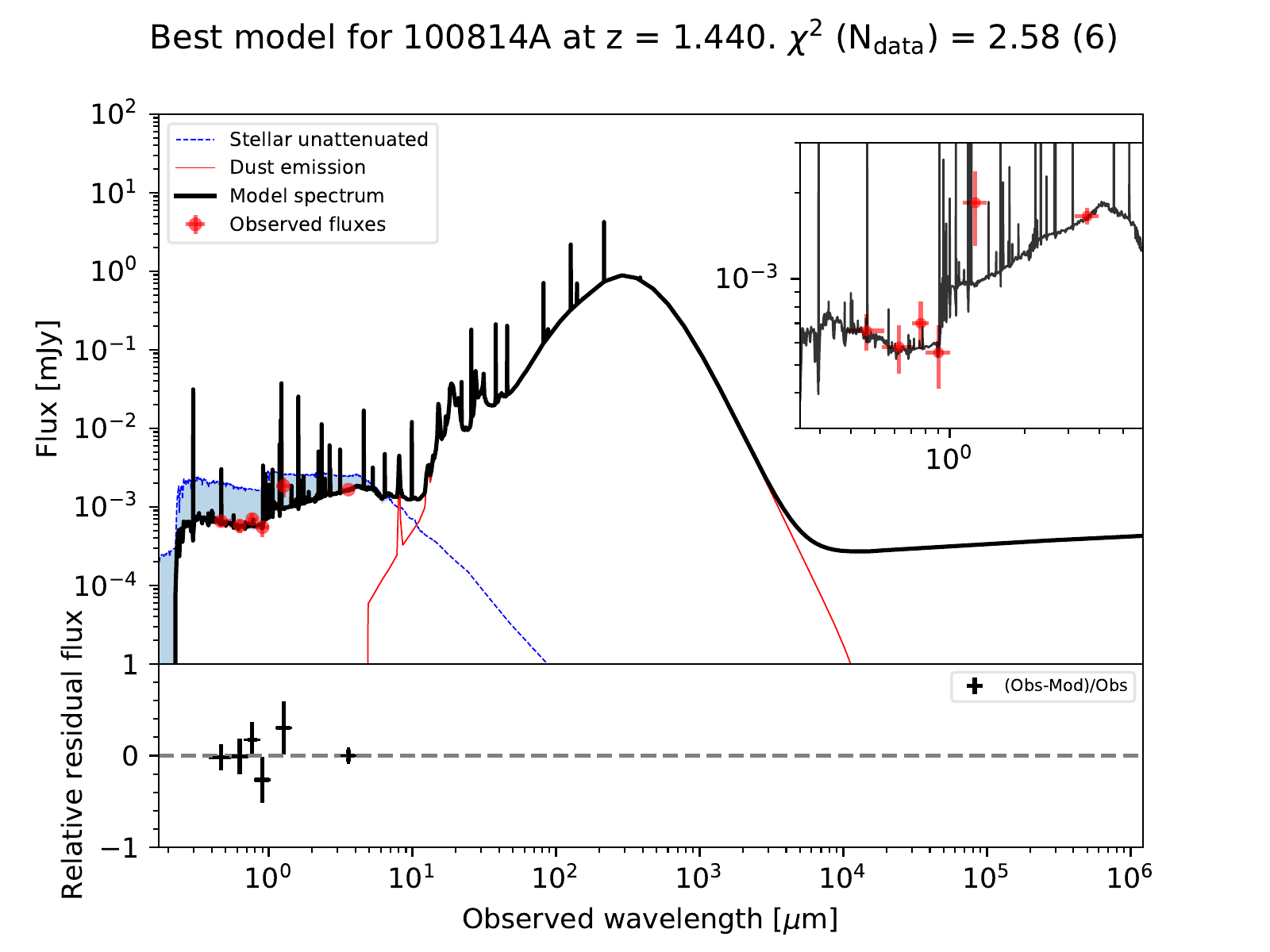}
	\caption{Best-fitted SED for GRBH100814A. Colour and symbols convention as in Figure \ref{best061121}.}
	\label{best100814}%
\end{figure}

The Balmer lines are measured by \cite{Kruehler15}, they derive a colour excess of $\rm E(B-V)_g = 0.08^{+0.26}_{-0.08}$ mag corresponding to an attenuation of $\rm A_V^{lines} = 0.24^{+0.8}_{-0.24}$ mag, and the lowest SFR of our sample with $\rm SFR_{H\alpha}$ = 3.2$^{+2.9}_{-0.7}$ $\rm M_\sun yr^{-1}$. The stellar mass  measured from the IRAC1 measurement in \cite{Perley16} is $\rm M_\star = 3.3 ~10^ 9 ~M_\sun$.\\ 
The best model is obtained for Z=0.008 and shown in Figure \ref{best100814}. No recent burst of star formation is added to the SFH. 
The SFR is equal to $\rm 4 \pm 2.3~ M_\sun yr^{-1}$ which is in good agreement with the one measured with the H$\alpha$ flux. The stellar mass is $\rm M_\star = 2.9 \pm 0.8 ~10^ 9 ~M_\sun$ in agreement with the one measured by \cite{Perley16}. We estimate a sSFR of $\rm 1.4 ~Gyr^{-1}$.\\
The dust attenuation is characterised by a colour excess $\rm E(B-V)_s = 0.05 \pm 0.04$ mag and a flat attenuation curve ($\delta=0.09 \pm 0.51 $). This results in an amount of visual extinction of $\rm A_{V}^{stars} = 0.35 \pm 0.4$ mag. No evidence for a UV bump is found.\\
A metallicity of Z=0.02 returned an equivalently good fit with very similar values for the SFR, the stellar mass, $\delta$ and $\rm A_{V}^{stars}$. \\
A recent burst in the SFH leads to an SFR estimation of $\rm 17.2 \pm 14.2 ~ M_\sun yr^{-1}$ and a colour excess of $0.16 \pm 0.07$ mag which are not in agreement with the SFR$\rm _{H\alpha}$ and the colour excess derived from the Balmer lines \cite{Kruehler15}. For this reason, we keep the results without any recent burst in the SFH as the reference.

\subsubsection{GRBH 120119A}
\begin{figure}
	\centering
	\includegraphics[width=\columnwidth]{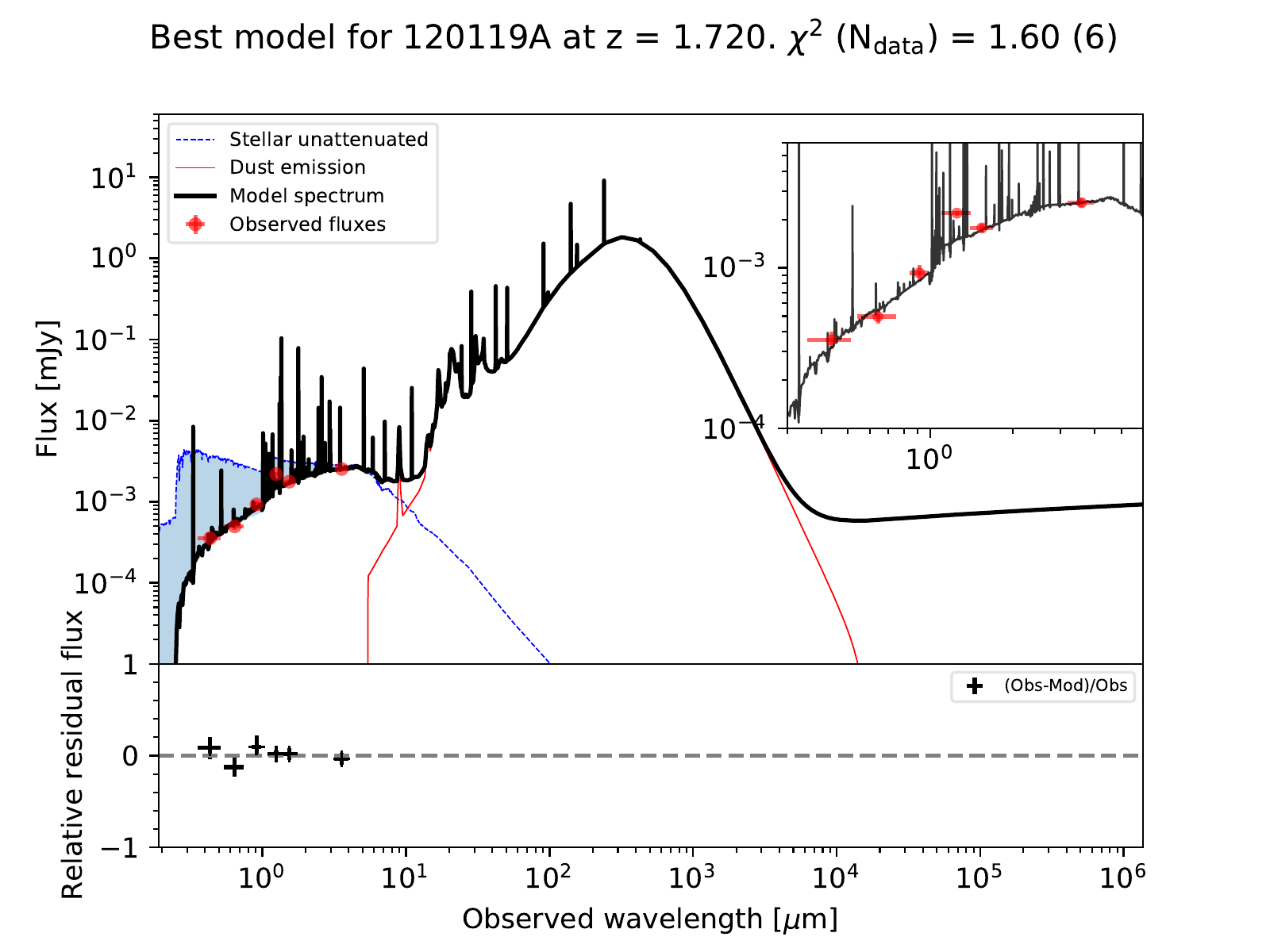}
	\caption{Best-fitted SED for GRBH120119A. Colour and symbols convention as in Figure \ref{best061121}.}
	\label{best120119}%
\end{figure}

The Balmer lines are measured by \cite{Kruehler15}, they derive a colour excess for the emission lines of $\rm E(B-V)_g=0.35^{+0.16}_{-0.14}$ mag corresponding to a visual attenuation of $\rm A_V^{lines} = 1.07^{+0.49}_{-0.43}$ mag, and a $\rm SFR_{H\alpha}$ of $\rm 43^{+24}_{-14}  ~M_\sun yr^{-1}$. The stellar mass  measured from the IRAC1 measurement in \cite{Perley16} is $\rm 8 ~10^9 ~M_\sun$.\\
The best model is obtained for Z=0.008 and shown in Figure \ref{best120119}. 
The SFR is found equal to $\rm 25.5 \pm 14.1~ M_\sun yr^{-1}$ marginally consistent with the one measured with the H$\alpha$ flux. A burst of star formation is found with an amplitude of $23 \pm 15 \%$ of the total mass and an age of $64 \pm 33$ Myrs. The stellar mass is $\rm M_\star = 3.8 \pm 1.2 ~10^ 9 ~M_\sun$  a factor of  2.1  lower than the one measured by \cite{Perley16}. We estimate a sSFR for this starbursting galaxy of $\rm 6.7 ~Gyr^{-1}$. The dust attenuation is characterised by a colour excess $\rm E(B-V)_s = 0.24 \pm 0.06$ mag and a attenuation curve steeper than the one of C00 ($\delta=-0.39 \pm 0.22 $). This results in an amount of visual extinction of $\rm A_{V}^{stars} = 0.74 \pm 0.28$ mag. No evidence for a UV bump is found. A metallicity of 0.02 returned a slightly worse $\chi^2$ with similar values for the SFR, stellar mass, $\delta$ and $\rm A_{V}^{stars}$. \\
\\
As for GRBH 080805, in order to improve the agreement between the SFR found from the SED and from H$\alpha$ measurements, we checked whether the presence of a very young starburst could enhance the current star formation rate. We ran again CIGALE with a recent burst taking place within the last 10 to 20 Myrs. The resulting fit is slightly better with a lower $\chi^2$. The SFR  increases to $\rm 44.4 \pm 26.6 ~ M_\sun yr^{-1}$, in very good agreement with the H$\alpha$ measurements. The burst of star formation is found with an amplitude of $16 \pm 15 \%$ of the total mass and an age of $15 \pm 4$ Myrs. The stellar mass is only reduced to $\rm M_\star = 3.6 \pm 1.4 ~ 10^9 ~M_\sun$. The attenuation curve becomes shallower with $\delta=-0.29 \pm 0.24$. The amount of attenuation is $\rm A_V^{stars}= 0.95 \pm 0.39 $ mag.  We keep the result with a burst age older than 20M yrs as the reference.\\
\begin{table*}
	\renewcommand{\arraystretch}{1.5}
	\caption{Results of the SED fitting.}             
	\label{tab:results}      
	\centering          
	\begin{tabular}{c c c c c c c c c c c}     
	\hline\hline       
	Ident & $z$ & SFR & M$_\star$ & $\rm A_V^{stars}$ & $\rm E(B-V)_s \over{E(B-V)_g}$ &  $\delta$ & sSFR & Recent burst \\
	GRBH   &         & M$_\sun$ yr$^{-1}$ &  $10^9$M$_\sun$ & mag & &  slope & Gyr$^{-1}$ & y/n \\
	\hline               
	061121 & 1.31 & 62.0 $\pm$ 25.7 & 6.9 $\pm$ 2.4 & 0.88 $\pm$ 0.30 & 0.28& 0.22 $\pm$ 0.2 & 9.0 $\pm$ 4.9 & y  \\
	070802   & 2.45 & 32.2 $\pm$ 17.8 & 3.7 $\pm$ 1.6 & 0.86 $\pm$ 0.35 & 0.71 &  -0.12 $\pm$ 0.30 &  8.7 $\pm$ 6.1 & y \\
	080605   & 1.64 & 44.9 $\pm$ 22.9 & 12.3 $\pm$ 4.2 & 0.41 $\pm$ 0.29 & 0.35 &  -0.08 $\pm$ 0.38 & 3.6 $\pm$ 2.2  & y \\
	080607    & 3.04 & 35.2 $\pm$13.9 & 27.5 $\pm$8.2 & 1.39 $\pm$0.29 & -- &  0.05 $\pm$0.10 & 1.3 $\pm$ 0.6 & n \\
	080805 & 1.5 &   20.6 $\pm$ 12 &  3.4 $\pm$1.7 & 1.06 $\pm$0.34 & 0.47 & -0.45 $\pm$0.24 & 6.1 $\pm$ 4.6  & y\\
	100814A & 1.44 & 4 $\pm$2.3 & 2.9 $\pm$0.8 & 0.35 $\pm$ 0.4 & 0.6 & 0.09 $\pm$ 0.51 & 1.4 $\pm$ 0.9 & n\\
	120119A &1.73 & 25.5 $\pm$14.1 & 3.8 $\pm$1.2 & 0.74 $\pm$0.28  & 0.68 & -0.39 $\pm$0.22 & 6.7 $\pm$ 4.3 & y \\
	\end{tabular}
\end{table*}

\begin{figure}
	\centering
	\includegraphics[width=0.49\columnwidth]{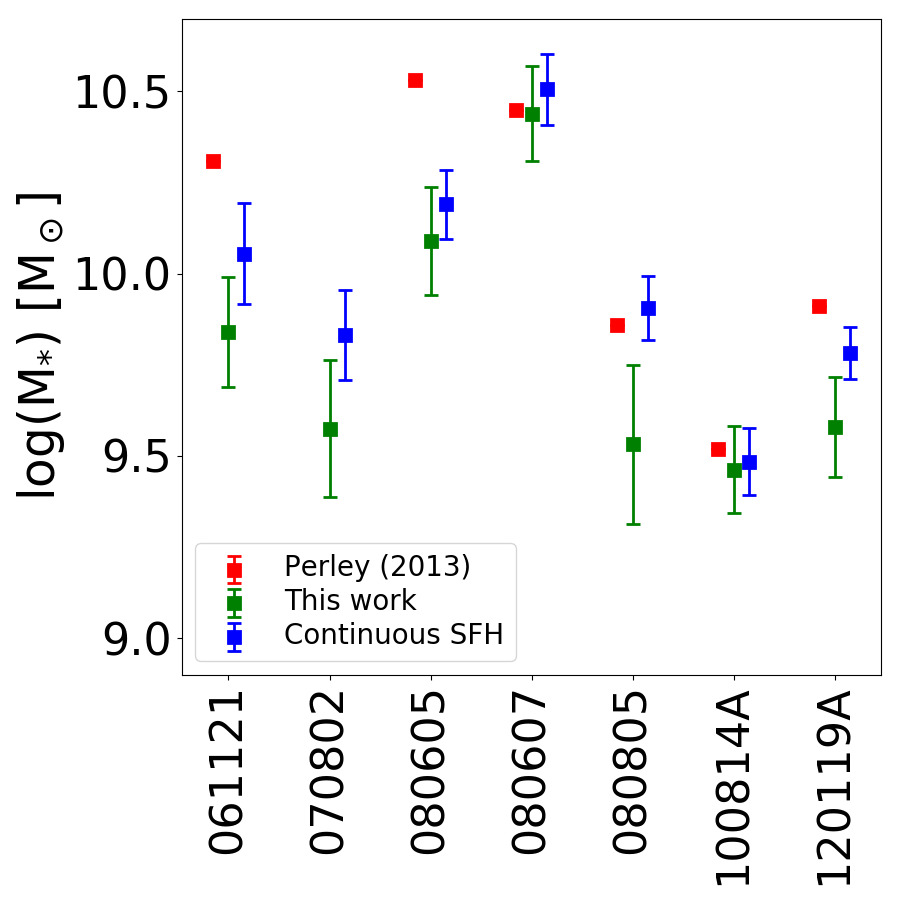}
	\includegraphics[width=0.49\columnwidth]{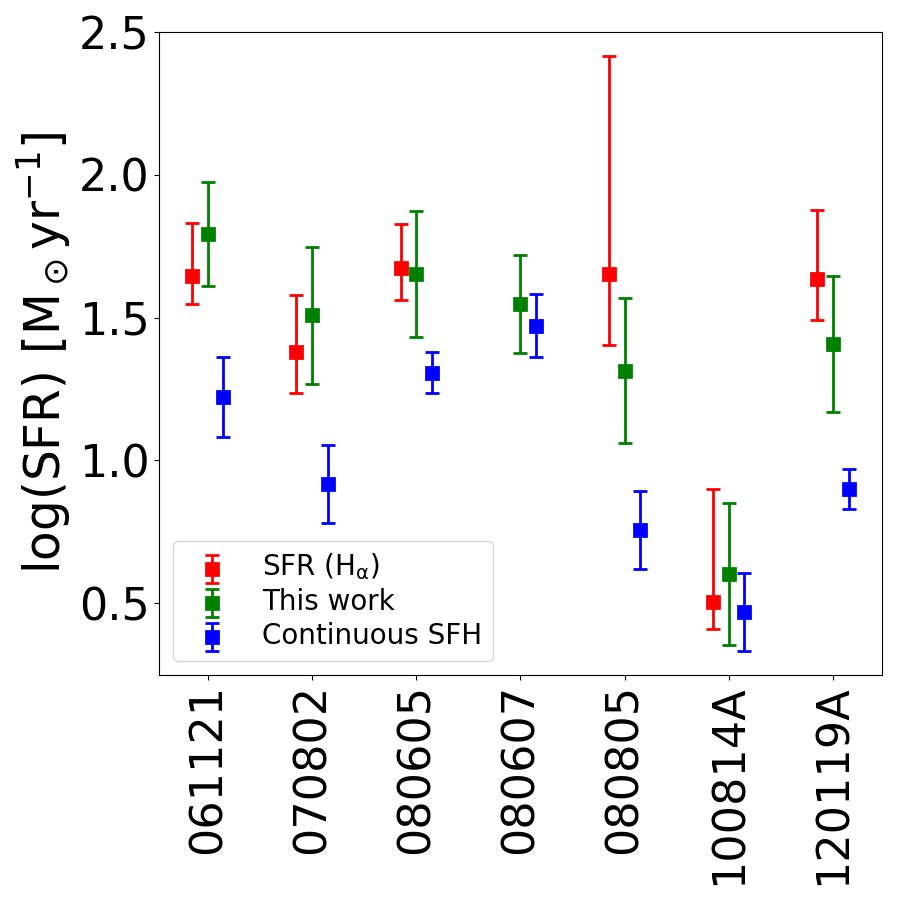}
	\caption{\textit{Left}: Stellar masses estimated from: (red) IRAC1 measurement in \cite{Perley16}, (green) this work with either a delayed or delayed+burst SFH, (blue) CIGALE using same configuration as in Table \ref{tab:parameter} but with a continuous SFH with a main stellar population age between 1 and 3.5 Gyrs. \textit{Right}: SFR estimated from: (red) $H_{\alpha}$ measurement from \cite{Kruehler15} and \cite{Vergani17}, (green) this work using either a delayed or delayed+burst SFH, (blue) CIGALE using same configuration as in Table \ref{tab:parameter} but with a continuous SFH with a main stellar population age between 1 and 3.5 Gyrs.}
	\label{fig:comparison_M_SFR}%
\end{figure}

\subsection{Star formation activity and SFR-M$_*$ comparison}

The results of the SED fitting are reported in Table \ref{tab:results}. The scenario with a recent burst on top of a delayed star formation history is favoured over a delayed only SFH for five of our seven GRB host galaxies. Generally, our stellar mass estimates are lower than in \cite{Perley16} where stellar masses are derived from the IRAC 3.6$\mu m$ luminosities using a continuous SFH. In order to check whether this difference is due to the SED fitting algorithm or to the adopted SFH we perform our fits with a constant SFR over 1 to 3.5 Gyrs similarly to the method in \cite{Perley16}. However, we are using all the photometry data and not only the IRAC1 detection, and the stellar population age was fixed according to the stellar mass in \cite{Perley16} which is not the case in our method. In the right panel of Figure \ref{fig:comparison_M_SFR} we compare the SFR derived with a continuous SFH and our results to the SFR$\rm _{H_{\alpha}}$ derived in \cite{Kruehler15}. The continuous SFH leads to lower SFR estimates that are not consistent with the SFR$\rm _{H_{\alpha}}$ except for GRBHs 080607 and 100814A where a recent burst is not required. On the contrary, the SFRs we derive with the delayed or delayed+burst SFH are always found in agreement with the SFR$\rm _{H_{\alpha}}$. As shown in the left panel of Figure \ref{fig:comparison_M_SFR}, the assumption of a constant SFR also impacts the mass measurements and our stellar mass estimates are  lower (up to a factor three) than the ones in \cite{Perley16} except for GRBHs 080607 and 100814A. The fact that we include all observations instead of just the IRAC1 measurement might partly explain the observed difference in some cases. In all cases the fits return a worse $\chi^2$ with a continuous SFH compared to delayed or delayed+burst SFH. We conclude that the discrepancy on the stellar mass estimation is mainly explained by the introduction of a strong and recent burst.\\
\begin{figure}
   \centering
	\includegraphics[width=9.5cm]{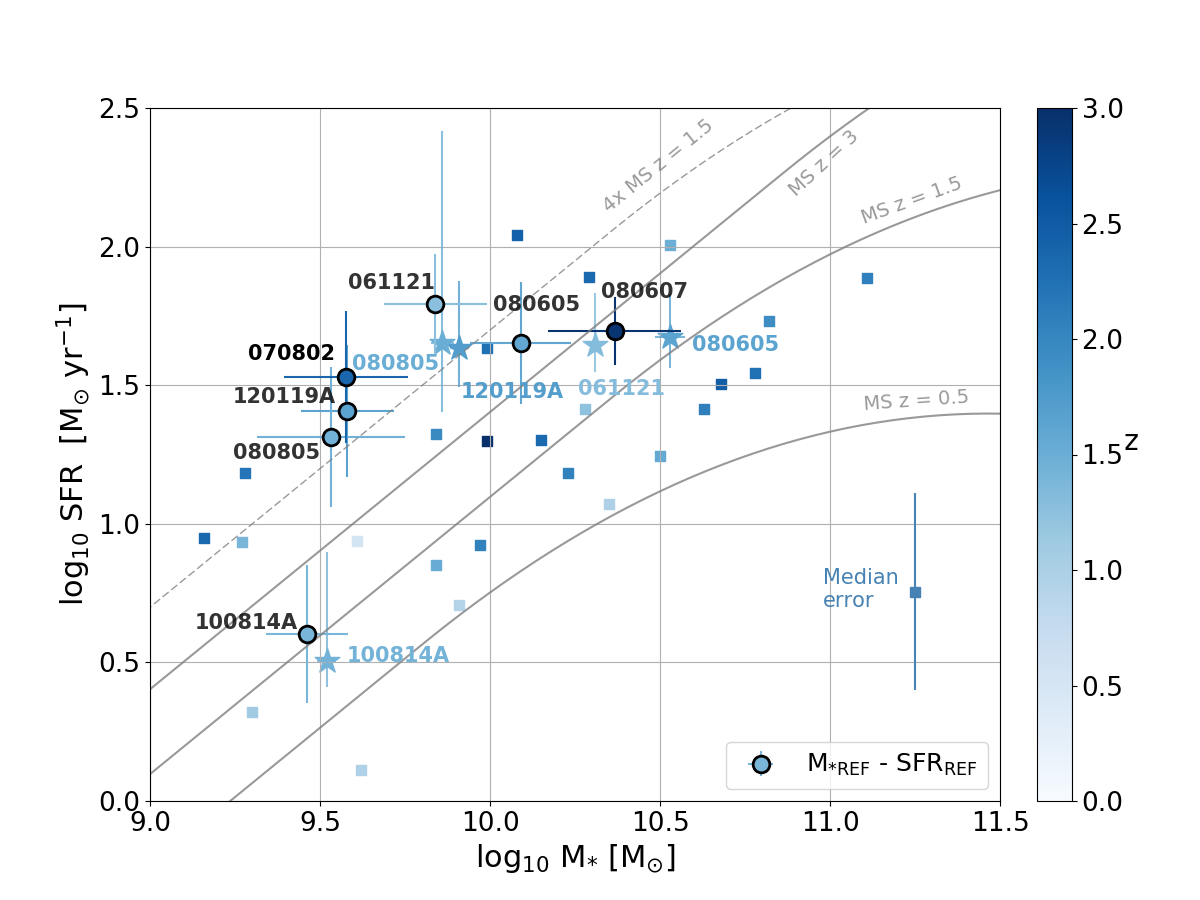}
	\caption{SFR-stellar mass relation. Blue points represent stellar masses from \cite{Perley16} and SFR from \cite{Kruehler15} where we highlighted the GRBHs in our sample using a star symbol and the corresponding name. The colour map encodes the redshift. The black circles with blue interiors correspond to the GRBHs of our sample with SFR and stellar masses derived from the SED fitting. The grey lines represent the main sequence relation from \cite{Schreiber15}, the dashed line is four times the main sequence at z=1.5 representative of starbursting galaxies (most of our our GRBHs are at z$\sim$1.5).}
    \label{fig:MS}%
\end{figure}
Figure \ref{fig:MS} represents the main sequence of all GRBHs having a stellar mass estimation in \cite{Perley16} and a SFR$\rm _{H_{\alpha}}$ in \cite{Kruehler15}. The main sequence of star-forming galaxies for z=0.5, 1.5 and 3, covering the GRBH redshifts, are represented using the relation from \cite{Schreiber15}. For the four GRBHs (061121, 080605, 080805, 120119A) with a delayed+burst SFH, our lower stellar mass estimates shift the galaxies to the starbursts domain  we define as four times the main sequence. These objects are dusty starbursting hosts and we do not expect such a strong effect for all GRB host galaxies, as illustrated by GRBH 100814A. 
Comparing our results with \cite{Kruehler11}, we find that our SFR estimations are consistent within errorbars for the 4 galaxies we have in common. We derive higher stellar mass for GRBH 080605 and GRBH 080607. They were limited to optical/NIR data up to the \textit{Ks} band, which explained the difference with our stellar mass estimation as we have a better NIR coverage. The old stellar population contributes significantly to the total stellar mass and dominates the NIR stellar emission, so the use of IRAC1 (and IRAC2 for GRBH 080607) detections allows a more reliable stellar mass estimation.\\
Given the importance of the stellar mass estimation in the evidence of a mild metallicity threshold \citep{Vergani15,Japelj16,Vergani17} necessary to form long GRB (LGRB), it would be of interest to revisit GRBH stellar masses estimation on a larger sample. LGRB hosts also offer the opportunity to identify and study distant low mass ($\rm \sim 10^8-10^{10} M_\sun$) starbursting galaxies generally absent or badly represented in blind surveys.\\

\begin{figure}
	\centering
	\includegraphics[scale=0.4]{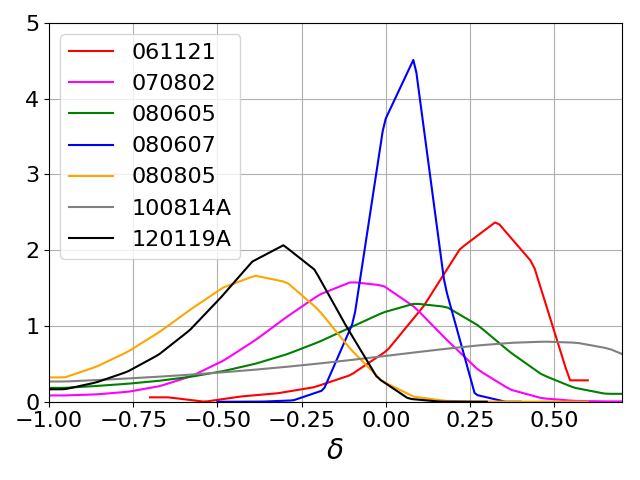}  
	\caption{Probability density functions of the attenuation slope for the GRBHs.}
	\label{fig:slope_pdf}%
\end{figure} 

\section{Dust attenuation in GRBHs}

In this section, we first characterise the attenuation curves we derived for the seven GRBHs of the Golden sample. Then we study the characteristics of the dust attenuation as a function of GRBH properties (the stellar mass, SFR, sSFR, and $A_V^{stars}$). We compare the attenuation curves derived from the SED fitting with the extinction curves along the GRB l.o.s. and finally compare these results to radiative transfer simulations to constrain dust-stars geometry in the ISM of these GRBHs.\\
  
\subsection{Measure of the attenuation curve slopes}

The dust attenuation is described using Equation \ref{eq:attlaw} where the colour excess, $\rm E(B-V)_s$, and the attenuation slope, $\delta$, are free parameters. The probability density functions obtained with CIGALE for the attenuation slope of each GRBH are shown in Figure \ref{fig:slope_pdf}. They are well defined with the exception of GRBH 100814A whose PDF is too flat due to the low amount of dust within the host galaxy characterised by a colour excess of $\rm E(B-V)_g = 0.08^{+0.26}_{-0.08}$ mag for the emission lines \citep{Kruehler15} and 
$\rm E(B-V)_s = 0.048^{+0.04}_{-0.04}$ mag for the stellar continuum. Therefore we will not consider this galaxy in the following.\\
There are a great variety of slopes with GRBH 070802, GRBH 080605 and GRBH 080607 similar to the Calzetti attenuation law defined by $\delta$=0. The attenuation curve of GRBH 061121 is flatter than C00. GRBH 080805 and GRBH 120119A have an attenuation curve slope as steep as the mean SMC extinction curve, corresponding to $\delta \sim -0.5$.
 
\subsection{Variation of the slope with GRB host properties}
\begin{figure}
	\centering
	\includegraphics[width=0.8\columnwidth]{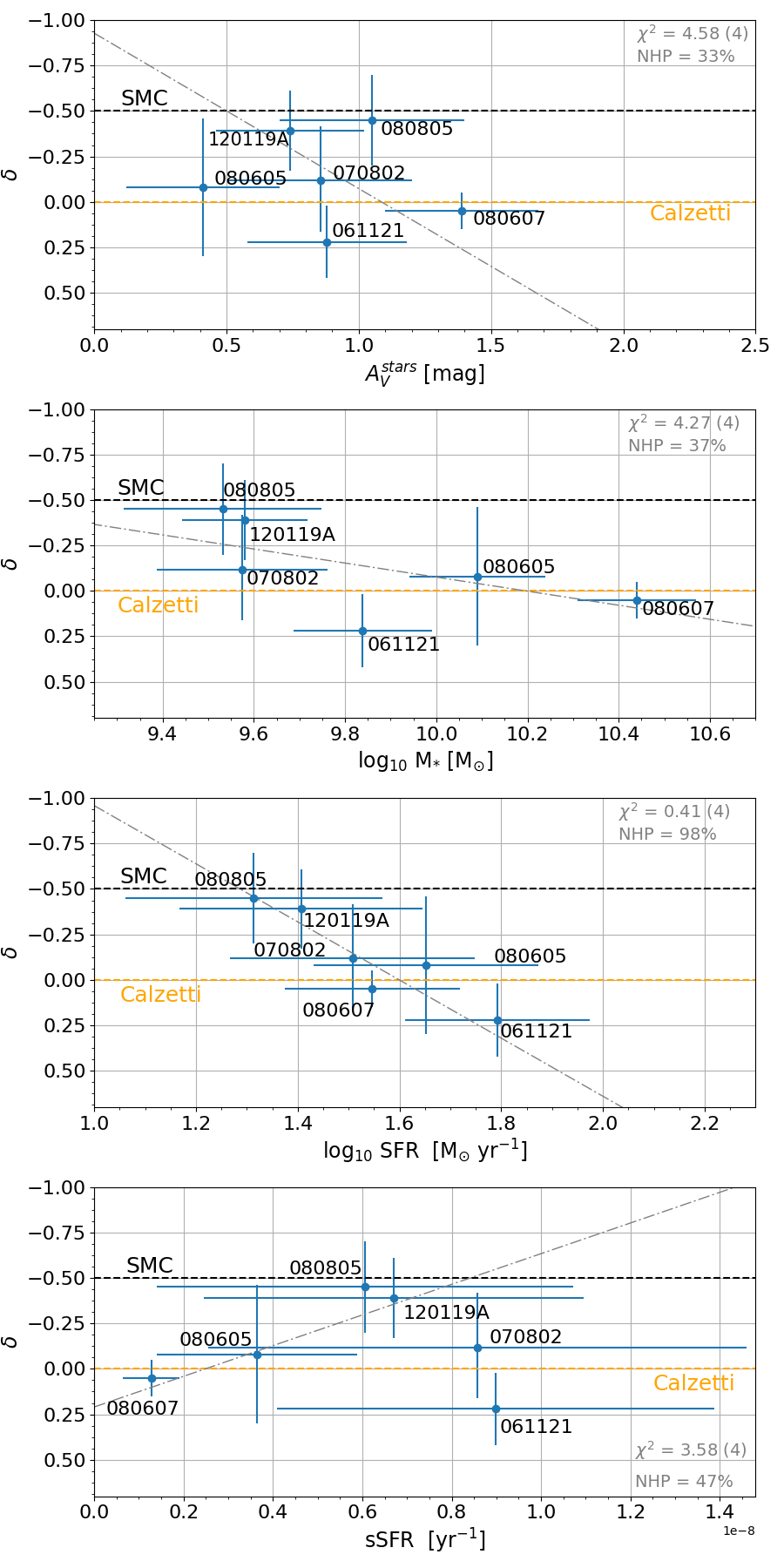}  
	\caption{\textit{Top}: Attenuation curve slope as a function of the SED continuum amount of attenuation. \textit{Middle top}: Attenuation curve slope versus the stellar mass. \textit{Middle bottom}: Attenuation curve slope versus the SFR. \textit{Bottom}: attenuation curve slope versus the specific SFR. The blue points correspond to the quantities derived with CIGALE. The black dashed line represents the slope of the mean SMC extinction curve and the yellow dashed line the slope of the Calzetti law. The grey dash-dotted line illustrates the best fit using an orthogonal distance regression (using SciPy's scipy.odr). The $\chi^2$, number of degrees of freedom and Null Hypothesis Probability (NHP) of these fits are indicated.}
	\label{fig:slope_Av_M_SFR_sSFR}%
\end{figure}

Figure \ref{fig:slope_Av_M_SFR_sSFR} represents the variation of the attenuation curve slope, $\delta$,  with some physical properties of the GRBH. We only have seven GRBHs in the sample so the conclusions are statistically limited. In the first panel representing the evolution of $\delta$ with $A_V^{stars}$, although no obvious trend is seen with a visual inspection of the plot, the orthogonal distance regression shows a trend with highly attenuated galaxies having flatter attenuation curves. This trend has already been observed for galaxies at z=1.5-3 \citep{Salmon16} and for galaxies at 0.5<z<2 \citep{Kriek13}. Radiative transfer simulations (i.e \citealt{Witt00, Chevallard13}) also predict a steeper dust attenuation curves at small attenuation optical depths, and shallow attenuation curve at larger optical depths. 
In the second panel, representing the variation of $\delta$ with the stellar mass, a clear trend is seen with lower (higher) stellar masses correlating with steeper (shallower) attenuation curves. This is also in agreement with the work of \cite{Salmon16} where the more massive galaxies in their spectroscopic redshift sample tend to prefer 'starbust' (i.e shallow) dust attenuation curves and less massive galaxies the SMC92 (i.e. steep) dust attenuation curves.\\
The third panel represents the variation of $\delta$ with the SFR and a clear trend is seen: GRBH with low (high) SFR have steeper (shallower) attenuation curves. \\
The bottom panel represents the evolution of $\delta$ with the sSFR and again although no trend can be seen with a visual inspection, the orthogonal distance regression shows that steep (shallow) attenuation curves are found in the more (less) active GRBHs. However the large uncertainties make this trend very uncertain. This result contradicts the results in \cite{Kriek13} which show that high  $\rm H_{\alpha}$ equivalent widths (i.e. high sSFRs) correlate with flat attenuation curves.

\subsection{Global dust attenuation and dust extinction along GRB l.o.s}
\begin{figure}
	\centering
	\includegraphics[width=0.8\columnwidth]{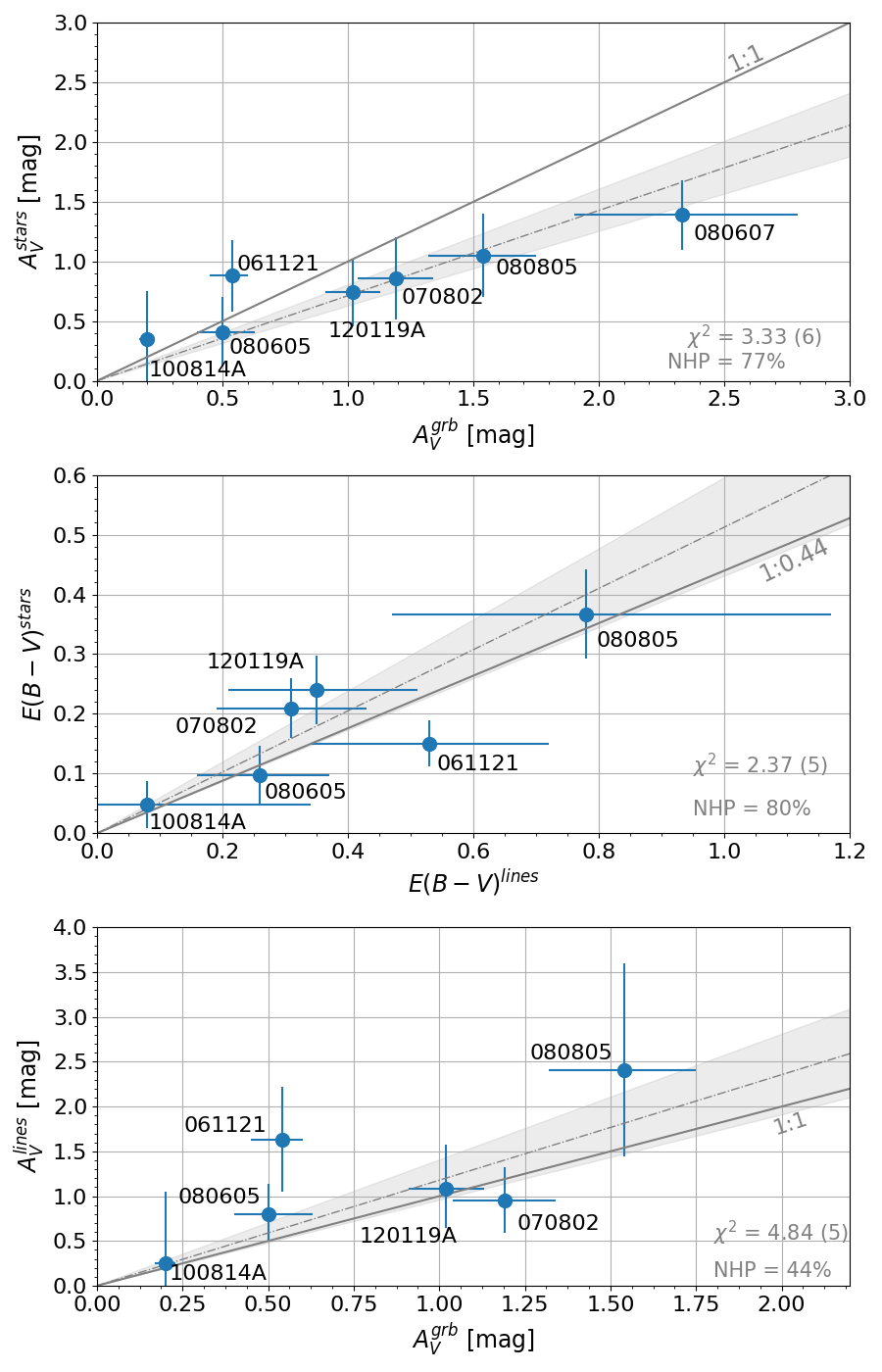}

	\caption{\textit{Top}: Amount of global attenuation in the GRBH in the V band as a function of the amount of extinction in the V band along the GRB l.o.s. \textit{Middle}: Colour excess for the stellar continuum as a function of the colour excess for the Balmer lines. The grey line represents the C00 relation. \textit{Bottom}: Amount of attenuation of the Balmer lines in the V band as a function as the amount of extinction in the V band along the GRB l.o.s. For all panels the grey dash-dotted line represents the best fit using an orthogonal distance regression (using SciPy's scipy.odr) and the grey shaded area its 1$\sigma$ uncertainty. The $\chi^2$, number of degrees of freedom and Null Hypothesis Probability (NHP) of these fist are indicated}
	\label{fig:Avgrb_Avstars}
\end{figure}
Top panel of Figure \ref{fig:Avgrb_Avstars} represents the evolution of the amount of dust extinction along the GRB l.o.s as a function of the amount of dust attenuation averaged over the whole GRBH. We see that they are correlated and GRB l.o.s are dustier than the average attenuation of the host galaxy. The orthogonal distance regression returns the relation $\rm A_{V}^{stars} = (0.71 \pm 0.11) \times A_{V}^{grb}$ for our sample. The LGRB progenitors are expected to be a result of the collapse of young massive stars and thus they should suffer more attenuation than the more evolved stellar population which had enough time to migrate outside their birth clouds (e.g. C00, \citealt{Charlot00, Granato00}). On a larger sample of 23 GRBHs selected with the criteria of having an obscured GRB l.o.s ($\rm A_V^{grb}$ > 1 mag), \cite{Perley13} find a similar trend: $\rm A_V^{stars} \lesssim A_V^{grb}$. However for $\rm A_V^{grb}$ < 1 mag, corresponding to less massive galaxies they find the opposite trend although dispersion is large. 
The middle panel of Figure \ref{fig:Avgrb_Avstars} represents the evolution of the colour excess for the stellar continuum as a function of the colour excess for the Balmer lines.  For our sample, we have $\rm E(B-V)_s=(0.51 \pm 0.08) \times E(B-V)_g$, which is in agreement with C00. The bottom panel of Figure \ref{fig:Avgrb_Avstars} represents the amount of attenuation for nebular lines in the V band as a function of the amount of extinction along the GRB l.o.s. The orthogonal distance regression returns: $A_V^{lines} = (1.18 \pm 0.22) \times A_V^{grb}$ which is consistent with the 1:1 relation within 1$\sigma$ uncertainty, meaning that the GRB and HII regions suffer a similar attenuation.

\subsection{Comparison of extinction and attenuation curves}
\begin{figure*}
	\centering
	\includegraphics[width=0.6\columnwidth]{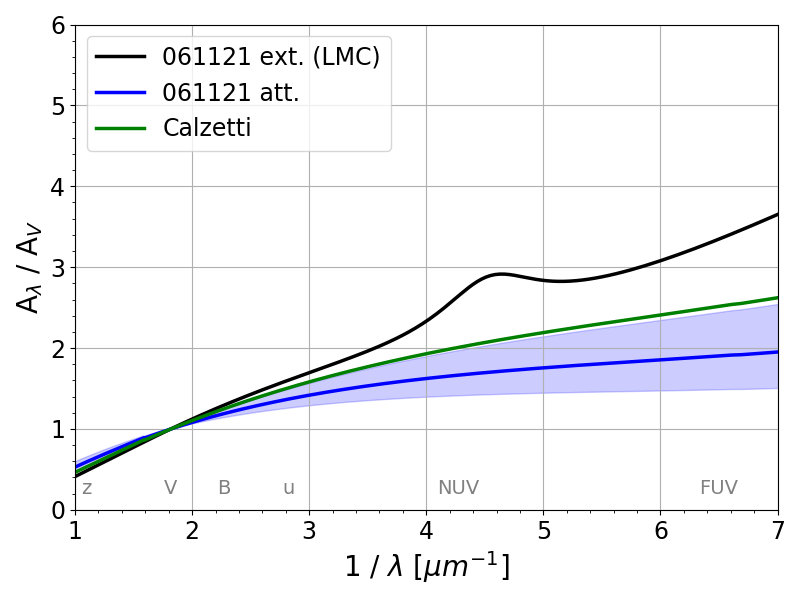}
	\includegraphics[width=0.6\columnwidth]{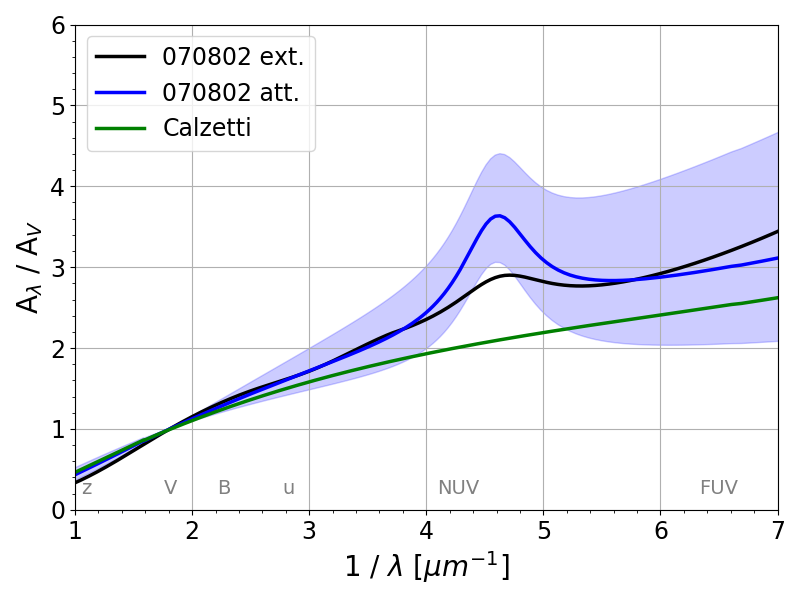}
	\includegraphics[width=0.6\columnwidth]{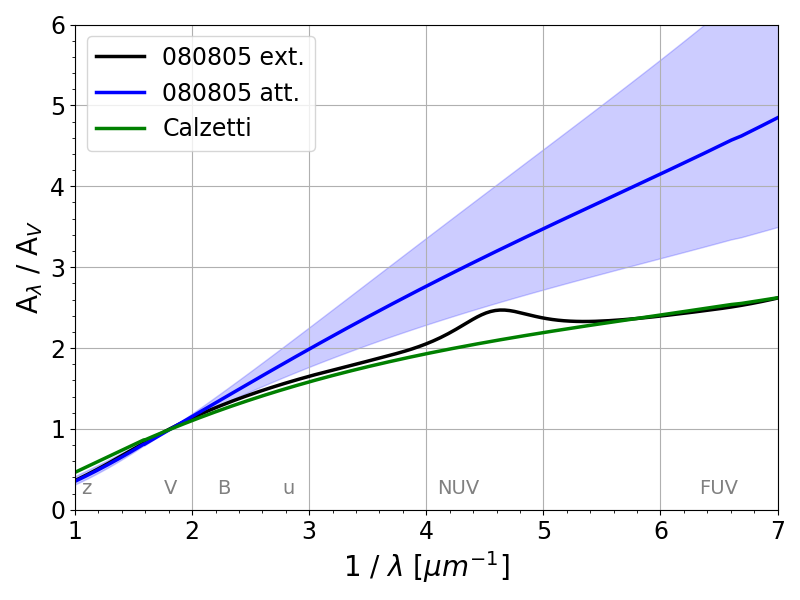}
	\includegraphics[width=0.6\columnwidth]{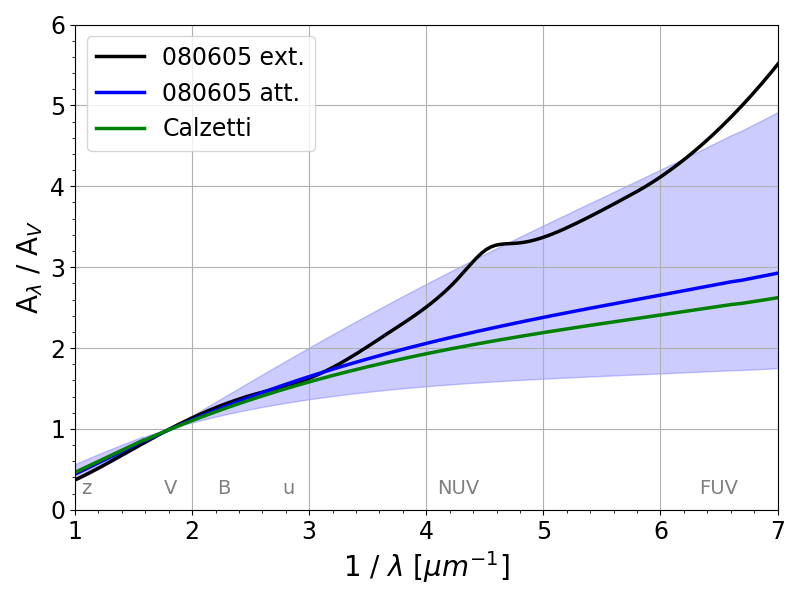}
    \includegraphics[width=0.6\columnwidth]{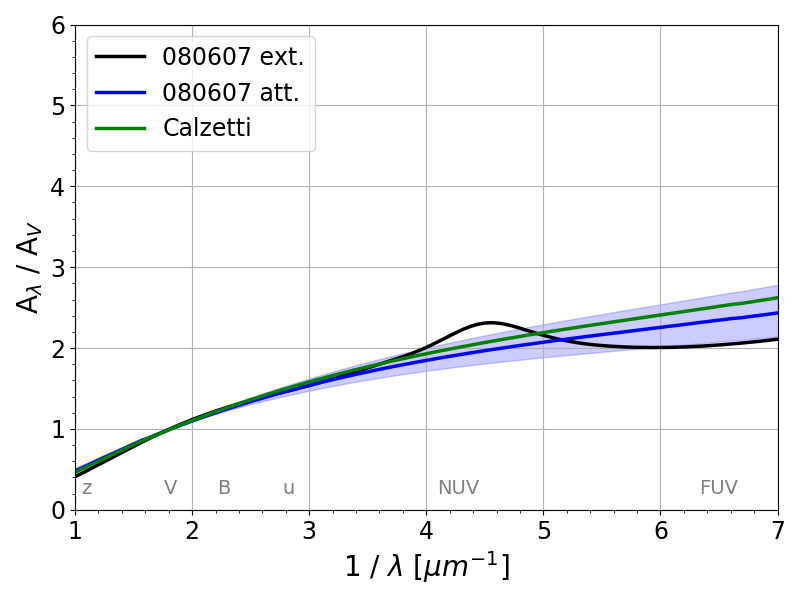}
	\includegraphics[width=0.6\columnwidth]{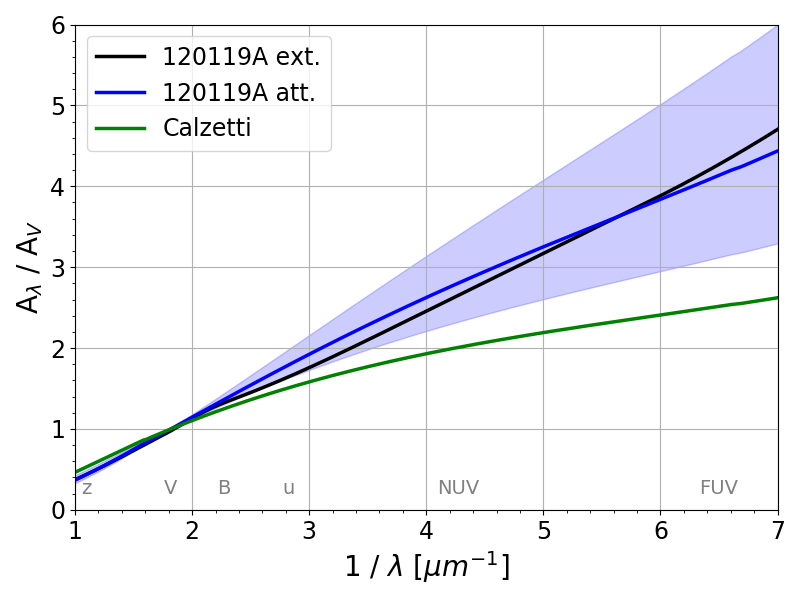}
	\caption{\textit{Top}: GRBH 061121, GRBH 070802 and GRBH 080805, \textit{Bottom}: GRBH 080605, GRBH 080607 and GRBH 120119. The solid black line is the extinction curve normalised to $\rm A^{grb}_{V}$, the blue one is the attenuation curve normalised to $\rm A^{stars}_{V}$ and the green one is the normalised Calzetti law. The blue shaded area represents the 1$\sigma$ uncertainty on the attenuation curve.}.
	\label{fig:ext_vs_att}%
\end{figure*}
The attenuation curves derived by the SED fitting are compared to the extinction curves along GRB l.o.s on Figure \ref{fig:ext_vs_att}. There is a great variety of cases that can be sorted into three categories. Category A, in which the attenuation curve is shallower than the extinction curve (061121 and 080605), category B, in which both attenuation and extinction curves are similar (080607, 070802, 120119A), and category C, in which the attenuation curve is steeper than the extinction curve (080805). The attenuation curves of 061121, 080605, and 080607 are consistent with the Calzetti law, those of 070802, 080805, and 120119A are not. We note than the better constrained attenuation curves are those with radio (GRBH 061121) and sub-millimeter (GRBH 080607) detections. 

\subsubsection{General shape}
Category A corresponds to the case with a flat attenuation curve and a steeper extinction curve. A far-UV rise in the extinction curve indicates the presence of small silicate grains and the UV bump the presence of PAHs or small carbonaceous grains. 
A shallow attenuation curve is usually obtained for high optical depth (\citealt{Witt00,Seon16}), highly inclined galaxies (\citealt{Pierini04,Inoue05}), a high density environment \citep{Inoue05} or a clumpy ISM (\citealt{Witt00,Seon16}). \\
\\
Category B corresponds to similar attenuation and extinction curves. At low optical depth an ISM in which dust and stars are uniformly mixed or stars surrounded by dust (shell geometry) with homogeneous and clumpy local dust distribution can reproduce an attenuation curve with a similar slope as the extinction curve, even as steep as the SMC extinction curve \citep{Witt00}. At higher optical depth only the shell geometry with an homogeneous dust distribution is able to produce such a steep attenuation curve from SMC dust type \citep{Witt00}.\\
The case of GRB/GRBH 080607 is interesting because it has the highest redshift in our sample and the extinction curve is unusually flat. This may be due either to variations of dust properties and/or geometry in the ISM. The geometrical effect is supported by the detection of a molecular cloud in the afterglow spectrum at >100 pc from the GRB, probably arising in a neighboring star-forming region \citep{Prochaska09}. 
Another possible explanation of the flat extinction curve is due to the dust composition: in the earliest evolutionary stage of dust evolution the dust content is dominated by large grains originated from stellar sources which results in a flat extinction curve ($\lesssim$ 1 Gyr). Later on, the abundance of small grains increases through shattering and accretion processes which results in the progressive apparition of a 2175\AA\ bump and a steeper extinction curve (\citealt{Asano13,Asano14}). This explanation can be supported by the redshift of GRBH08607 of z=3.04 which corresponds to an age of the Universe of about 2.1 Gyrs adopting the $\rm \Lambda CDM $ cosmology of \cite{Planck15}. The rather high value of $\rm R_V\sim$ 4 measured from the extinction curve is also indicative of a large grain size distribution though the uncertainty is quite large (\citealt{Prochaska09,Zafar11}). The forthcoming SVOM mission might potentially bring more data to constrain the dust evolution processes at high redshift.\\
\\
Category C corresponds to an attenuation curve steeper than the extinction curve. A steep attenuation curve can be the result of a low optical depth (\citealt{Witt00,Seon16}) for example in a face-on galaxy. It may also result from an age-dependent attenuation where the young stellar population is embedded in dense clouds and more attenuated than the old stellar population which had time to migrate to the more diffuse ISM \citep[e.g.][]{Inoue05}.\\
\begin{table*}
	\renewcommand{\arraystretch}{1.5}
	\caption{Summary of the comparison with radiative transfer model of \cite{Seon16}. }
	\label{comp_rad}
	\centering
	\begin{tabular}{ccc}
	\hline
	 & Dust model & ISM structure \\
	\hline
	\hline
    061121 & WD-LMC & Dusty + Very clumpy\\
    070802 & WD-LMC & Dusty + Very clumpy\\
    080605 & WD-LMC & Dusty + Very clumpy\\
    080607 & WD-MW & Shell + Clumpy\\
    080805 & WD-MW & Dusty + Homogeneous or Shell + Clumpy\\
    120119A & WG-SMC & Dusty + Homogeneous or Shell + Clumpy\\
	\end{tabular}
\end{table*}

\subsubsection{Presence of the UV bump in the attenuation curves}

For many years, there has been a debate whether a Calzetti-like attenuation curve can ever arise if the underlying extinction curve exhibits a UV bump. While the works of \cite{Gordon97, Witt00} and \cite{Pierini04} suggested that a Calzetti-like attenuation law implies a bump-free extinction curve, for example, a SMC-like law, \cite{Silva98, Granato00} could reproduce a Calzetti-like attenuation law with an underlying MW extinction curve. In our sample of GRB extinction curves, five out of six exhibit a UV bump with four spectroscopically confirmed detections. The presence or lack of a UV bump in the attenuation curves should provide valuable information. \\
The UV bump is best constrained using spectroscopic measurements of galaxies \citep[e.g.][]{Noll07} but intermediate filter bands also allow to constrain the bump amplitude \citep[e.g.][]{Buat11, Kriek13}. However, this is very difficult using broad filter bands only as it relies only on one filter band overlapping the UV bump with a passband width higher than the bump width. In our sample, GRBHs 061121, 070802, 080605, 080805 and 120119A have a filter band which overlaps the UV bump. The rest-frame spectral coverage of each GRBH observation can be seen in Appendix C. For GRBH 070802, we already reported a bump amplitude of $\rm E_b = 5.3 \pm 2.6$ in Section \ref{sec:070802}. 
For the other GRBHs, we already ran CIGALE with the same configuration as described in Section \ref{sec:SED_fitting} with a varying UV bump amplitude ($E_b>0$) and did not find clear evidence of its presence. The PDFs of the UV bump amplitude exhibit a simple power law shape with the highest probability associated to an amplitude of zero and a decreasing probability as the amplitude increases. Given the rather weak UV bump amplitudes measured in the extinction curves, we run CIGALE again letting the UV bump amplitude vary from -2 to 2 this time, similarly as in \cite{Buat12}, in order to obtain better defined PDFs for a more reliable Bayesian estimation of the UV bump amplitude if PDFs are centred around zero. The results are reported in Table \ref{tab:UV_bump}, all UV bump amplitudes are compatible with zero within 1$\sigma$ uncertainty. The estimation of the other  parameters remain unchanged. We conclude that we do not find evidence of a UV bump with the possible exception of GRBH 070802, but we can not exclude the presence of a weak UV bump in the attenuation curves. Further observations using spectroscopy or intermediate filter bands are required in order to have a better constraint.
\begin{table}[h!]
	\renewcommand{\arraystretch}{1.2}
	\caption{Constraints on the UV bump amplitude when allowed to vary between -2 and 2 in the CIGALE run.}
	\label{tab:UV_bump}
	\centering
	\begin{tabular}{cccc}
	\hline
	GRBH  & Filter band  & $E_{b, ext.}$ & $E_{b, att.}$ \\
	 & overlapping & $=c_3 / \gamma^2$ &\\
	 & UV bump & &\\
	\hline
	\hline
    070802  & yes & 1.49 $\pm$ 0.30 & 5.3 $\pm$ 2.6 \\
    080605 &yes & 1.19 $\pm$ 0.39 & -0.26 $\pm$ 1.12 \\
    080607 & no & 2.05 $\pm$ 0.48 & --\\
    080805 & yes & 1.33 $\pm$ 0.70& -0.16 $\pm$ 1.14\\
    120119A & yes & none & 0.69 $\pm$ 0.97\\
    \hline
    061121$^{(a)}$ & yes & no spectro. & 0.03 $\pm$ 0.96 \\
	\end{tabular}
	\tablefoot{$(a)$: GRB 061121 is the only afterglow for which the extinction curve is derived with photometric data only.}
\end{table}

\subsection{Comparison with radiative transfer simulations}
\label{sec:RT}
\begin{figure*}
	\centering
	\includegraphics[width=0.6\columnwidth]{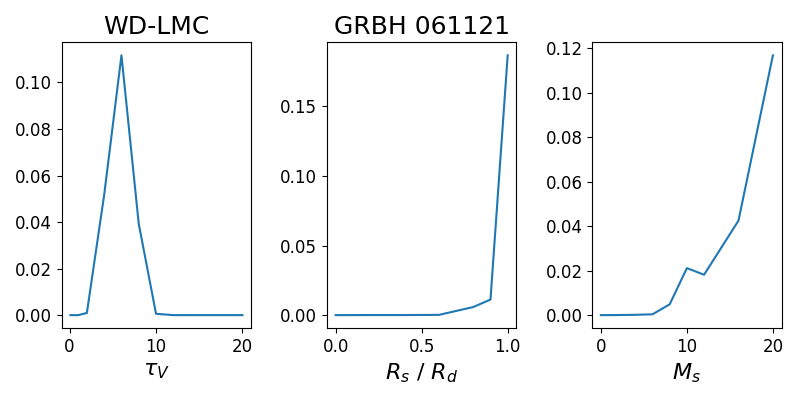}
	\vrule width 0.5mm height 3cm
    	\includegraphics[width=0.6\columnwidth]{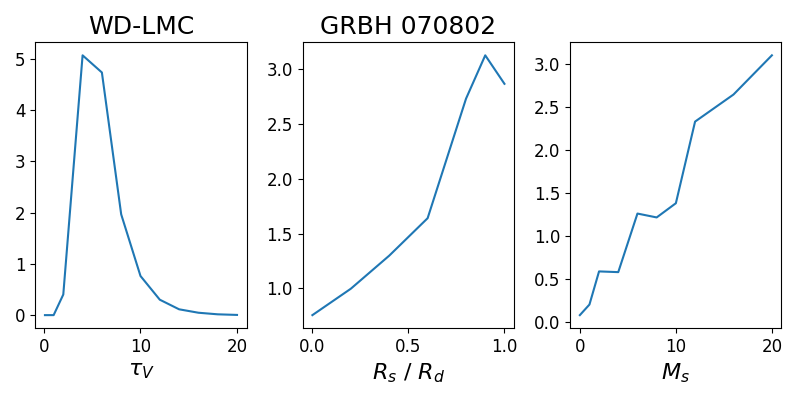}
    	\vrule width 0.5mm height 3cm
	\includegraphics[width=0.6\columnwidth]{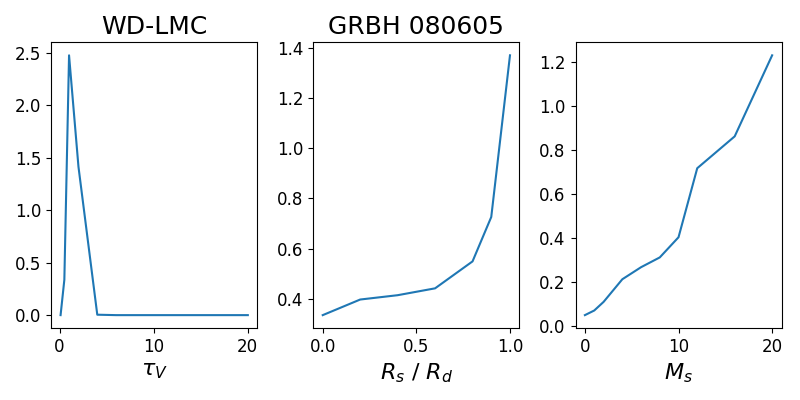}
	\includegraphics[width=0.6\columnwidth]{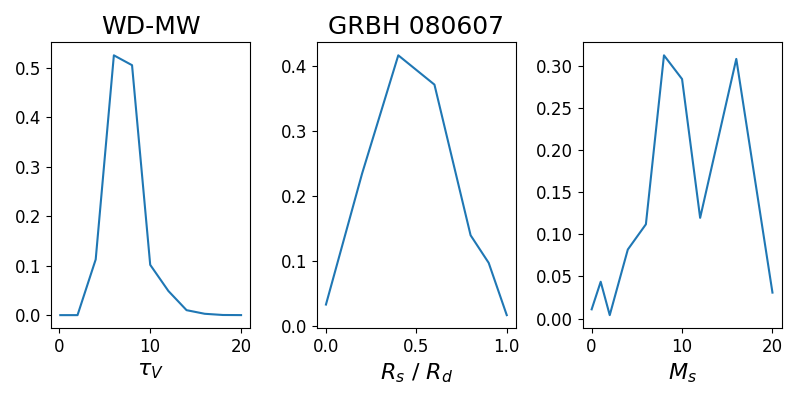}
	\vrule width 0.5mm height 3cm
	\includegraphics[width=0.6\columnwidth]{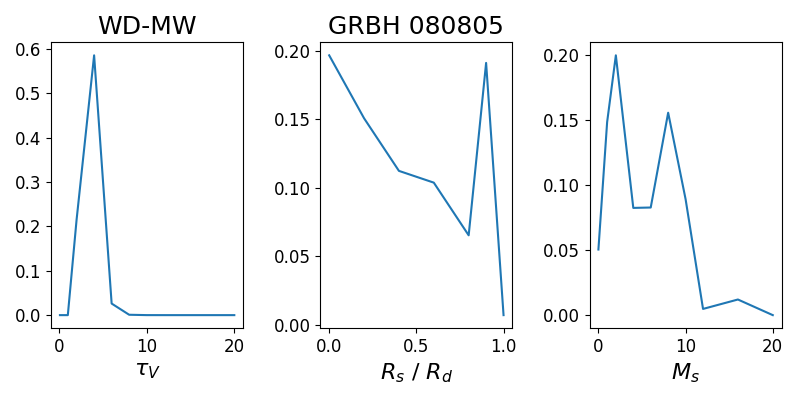}
	\vrule width 0.5mm height 3cm
	\includegraphics[width=0.6\columnwidth]{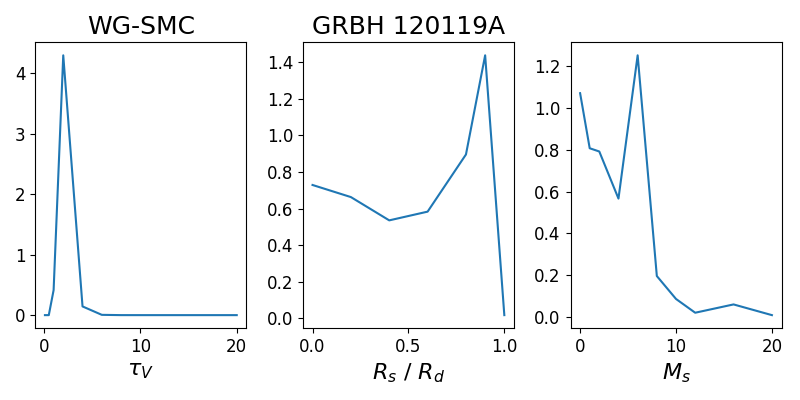}
	\caption{Probability density functions of $\tau_V$, $R_s/R_d$ and $M_s$ for each GRBH with the chosen dust model.}
	\label{fig:RT_pdf}%
\end{figure*}

To investigate the ISM geometry, we compared the derived attenuation curves with CIGALE with the results of the recent radiative transfer code developed by \cite{Seon16}. This model offers the possibility to choose between three theoretical dust models adopted from \cite{WD01} and \cite{Draine03} for the MW, LMC, and SMC-bar, and two empirical models for the MW and SMC from \cite{Witt00}. The former dust models are noted WD-MW, WD-LMC and WD-SMC and the later WG-LMC and WG-SMC. \\
We selected the dust model according to the GRB extinction curves (cf. Table \ref{comp_rad}). Following \cite{Schady12} GRB 061121 is best fitted with a LMC-like extinction curve\footnotemark\ . The extinction curve of GRB 070802 is very similar to the mean LMC2 super shell extinction curve \citep{Gordon03} so we choose a LMC dust model. The far-UV rise of the extinction curve of GRB 080605 is similar to an SMC extinction curve, however due to the spectroscopically confirmed detection of a UV bump, we chose an LMC dust model. For GRB 080607 and GRB 080805 due to the flat FUV part and the presence of a UV bump, we select a MW dust model. For GRB 120119A, we choose an SMC-like dust model according to Zafar et al. (2018, submitted). For SMC and MW like extinction curves both theoretical and empirical dust models are tested and we kept the one reproducing the best the attenuation curve derived with CIGALE.
\footnotetext{We note that \cite{Covino13} derived an SMC extinction curve for GRB 061121. We have tested both WG-SMC and WD-SMC dust model and obtained similar results.}
\\
The attenuation curves computed with the radiative transfer model of \cite{Seon16} are defined with three parameters: $\tau_V$, $R_s/R_d$ and $M_s$. $\tau_V$ is the optical depth in V band and take discrete values between 0.1 and 20. $R_s/R_d$ corresponds to the ratio of the distribution radius of stars and dust respectively, it takes discrete values between 0 and 1. $R_s/R_d$=0 corresponds to a compact OB association surrounded by a cloud and $R_s/R_d$=1 corresponds to the uniform distribution of photon sources over the spherical dusty medium (similar to the 'DUSTY' configuration of \citealt{Witt00}). To qualitatively assess our results we followed a formalism similar to the one of \cite{Witt00} by referring to a Dusty ISM for $R_s/R_d \geq$0.8 when stars are approximately uniformly distributed with dust and to a Shell ISM for $R_s/R_d \leq$0.5 with the stars surrounded by a dust cloud. $M_s$ is the Mach number governing the variance of the ISM density field , in other words, quantifying the clumpiness of the ISM. It takes discrete values between 0 and 20. $M_s=0$ corresponds to a homogeneous ISM and $M_s=20$ to a very clumpy ISM. In the following, we will refer to a homogeneous ISM for $M_s\leq$3, to a Clumpy ISM for 4$\leq M_s \le$15 and to a Very Clumpy ISM for $M_s\geq$15.\\
We compared the attenuation curves derived with CIGALE to the attenuation curves computed with the radiative transfer model for the different values of $\tau_V$, $\frac{R_s}{R_d}$, $M_s$. For each attenuation curve derived with CIGALE, a $\chi^2$ grid is then built using
\begin{ceqn}
\begin{equation}
\chi^2(\tau_V,\frac{R_s}{R_d},M_s) = \sum_{i=1}^{N_{data}} \left( \frac{A_{\lambda, i}^{RT} - A_{\lambda, i}^{SED}}{\sigma_{i}^{SED}} \right)^2
\end{equation}
\end{ceqn}
with $A_{\lambda, i}^{RT}$ the amount of attenuation from the radiative transfer model of \cite{Seon16}, $A_{\lambda, i}^{SED}$ and $\sigma_{i}^{SED}$ the amount of attenuation and its 1$\sigma$ uncertainty derived with CIGALE. These quantities are evaluated at the $N_{data}$ wavelengths used by \cite{Seon16} and covering the GRB Hosts observations.\\
From the $\chi^2$ grid, we compute the PDF of each parameter by marginalising the PDF over the two other parameters. The PDFs are shown in Figure \ref{fig:RT_pdf} and the results are summarised in Table \ref{comp_rad}. The cases of GRBH 061121, 070802, 080605 are similar with a low optical depth and high values for $R_s/R_d$ and $M_s$ indicating that a Dusty and Very clumpy ISM best reproduces the attenuation curves derived with CIGALE. Simulations with $R_s/R_d$>1 are not provided in the publicly available simulations of \cite{Seon16}\footnotemark\ .  It would be interesting to perform radiation transfer simulations with values of $R_s/R_d$ higher than 1 as in the work of \cite{Witt00} where the \textit{CLOUDY} configuration has $R_s/R_d \sim$1.45. For GRBH 080607, the $R_s/R_d$ PDF peaks at $\sim 0.5$ and the $M_s$ PDF disfavours a homogeneous ISM comparing to a Clumpy or even Very clumpy ISM. For GRBH 080805 and 120119A, both having steep attenuation curves, a bimodal solution is found with either a Dusty-homogeneous ISM or a Shell-Clumpy ISM. The attenuation curves computed with the radiative transfer model best fitting the attenuation curves derived with CIGALE are represented in Figure \ref{fig:RT_best_fit}. All attenuation curves computed with the radiative transfer model with a LMC or MW dust grain type that best fit the CIGALE attenuation curves exhibit a UV bump but the uncertainty associated to the CIGALE attenuation curves does not allow to constrain them.
\footnotetext{https://seoncafe.github.io/MoCafe.html}

\section{Conclusion}
We have selected 30 GRBs with a rest-frame coverage from UV to optical and find that the type of GRB extinction curves and the amount of extinction along the GRB l.o.s correlate with the host galaxies properties with the more extinguished GRBs preferentially found in the more massive hosts. The UV bump is also preferentially found in the most extinguished GRB l.o.s. It follows that without a rapid follow-up there is an observational bias towards SMC-like extinction curves for GRBs. This correlation presumably indicates that the GRB afterglow probes along its l.o.s not only the dust in the vicinity of the burst but also the dust located at large distances to be representative enough of the host galaxy.\\
From these 30 GRBs we have selected seven host galaxies whose SEDs have photometric measurements of the UV rest-frame emission and a UV-to-NIR spectrum reasonably well sampled (with at least six bands). The results of our SED fitting show that a recent burst of star formation is required for five out of seven GRBHs to recover the SFR measured from $\rm H_{\alpha}$ flux. This leads to a lower estimate of the stellar masses than the ones derived in \cite{Perley16} using a mass-to-light relation with the IRAC 3.6$\mu m$ luminosities. Consequently these GRBHs are shifted to the starburst domain, defined as four times the star-forming main sequence. For these relatively dusty GRBHs, we find that GRB and HII regions suffer a similar attenuation. Our estimates of the amount of attenuation for the whole galaxy, derived from the SED fitting, correlate with the amount of extinction along the GRB l.o.s., with the galaxy attenuation amounting to about 70\% of the GRB l.o.s. extinction. The ratio of colour excess between the stellar continuum and the Balmer lines is consistent with the one of C00. \\
The attenuation curves derived from the SED fitting show a great variety of cases: the attenuation curves have flatter, similar or steeper slopes than the extinction curves. Half of the sample is consistent with the C00 law and there is evidence of a UV bump in only one attenuation curve.
We find that steeper (shallower) attenuation curve slopes are found preferentially in the less (more) massive galaxies and with higher (lower) SFR and $A_V^{stars}$. These trends encourage us to further study these possible correlations on a larger sample of GRBHs.\\
Comparing our results with the radiative transfer simulations in \cite{Seon16}, we find a variety of dust-stars geometries. For half of the sample, a uniform distribution of stars and dust with a very clumpy local dust distribution reproduces the derived attenuation curves the best. In one case, the ISM structure is best represented with stars surrounded by a dust cloud with a clumpy local dust distribution. For the two steepest attenuation curves, two configurations are found almost as likely, either a uniform distribution of stars and dust with a homogeneous local dust distribution or stars surrounded by a dust cloud with a clumpy local dust distribution.\\
This method is promising and a dedicated radiative transfer simulation on each of these objects would allow to understand more precisely the dust-stars distribution in these GRBHs. \\

\vspace{3cm}
\paragraph{\textit{Acknowledgements}}
DC acknowledges H. Hirashita, A.K. Inoue, T.T. Takeuchi and D. Donevski for helpful and stimulating discussions. DC acknowledges support by the Centre National d'Etudes Spatiales and support by the Région Provence-Alpes-Côte d'Azur for the funding of his PhD. JJ acknowledges support from NOVA and NWO-FAPESP grant for advanced instrumentation in astronomy. This work was supported by the Programme National de Cosmologie et Galaxies (PNCG). SDV, SBo, RS, JP, acknowledges the support of the French National Research Agency (ANR) under contract ANR-16-CE31-0003 BEaPro (P.I.: SDV).

\nocite{*}
\bibliographystyle{aa}
\bibliography{aa}

\newpage
\onecolumn

\begin{appendix}
\section{GRB samples}

\begin{table}[htbp]
		\resizebox{0.91\textwidth}{!}{\begin{minipage}{\textwidth}
	\centering
	\caption{\label{tab:spec_sample} GRB for which the extinction curves are  derived by fitting with the FM formula in \cite{Zafar11,Zafar12}. The GRBs in bold are the ones in our Golden sample.}
	\renewcommand{\arraystretch}{2}
	\rotatebox{90}{
	\begin{tabular}{ccccccccccccc}
	\hline\hline
	GRB & z & $\rm T_{SED}$ & $\rm A_{V}^{Gal}$ & ext. curve & pow/bknp & $\rm A_{V}^{grb}$ & log$_{10}$ M$_{\odot}$ & $\rm A_{V}^{lines}$  & SFR$\rm _{H_{\alpha}}$ & $Z$ & References \\
	 & & hr & mag & & & mag & & mag & $M_{\odot}$ yr$^{-1}$ & 12 + log(O/H) & \\
	\hline
	050401 & 2.8983 & 14.7 & 0.172 & SMC-like & bknp & $0.65^{+0.04}_{-0.04}$ & 9.61 & & & & (1), (5)\\
	060729 & 0.5428 & 13.2 & 0.142 & SMC-like & pow & $0.07^{+0.02}_{-0.02}$ & 8.31 & $2.19^{+1.32}_{-1.45}$ & $0.96^{+2.21}_{-0.69}$ & & (1), (5), (6)\\
	070125 & 1.5471 & 31.5 & 0.139 & SMC-like & bknp & $0.30^{+0.04}_{-0.04}$ & & & & & (1)\\
	\textbf{070802} & \textbf{2.4541} & \textbf{2.00} & \textbf{0.071} & \textbf{LMC-like} & \textbf{pow} & $\mathbf{1.19^{+0.15}_{-0.15}}$ & & $\mathbf{0.95^{+0.37}_{-0.37}}$ & $\mathbf{24^{+11}_{-8}}$ & & (1), (6)\\
	071020 & 2.1462 & 5.12 & 0.160 & SMC-like & bknp & $0.43^{+0.04}_{-0.04}$ & $<$9.43 & & & & (1), (5)\\
	\textbf{080605} & \textbf{1.6403} & \textbf{1.74} & \textbf{0.360} & \textbf{Steep+bump} & \textbf{pow} & $\mathbf{0.50^{+0.13}_{-0.10}}$ & \textbf{10.53} & $\mathbf{0.80^{+0.34}_{-0.31}}$ & $\mathbf{47^{+17}_{-12}}$ & $\mathbf{8.54^{+0.09}_{-0.09}}$ & (2), (5), (6) \\
	\textbf{080607} & \textbf{3.0368} & \textbf{0.08} & \textbf{0.185} & \textbf{Flat+bump} & \textbf{pow} & $\mathbf{2.33^{+0.46}_{-0.43}}$ & & \textbf{10.45} & & & (1), (5)\\
	\textbf{080805} & \textbf{1.5042} & \textbf{1.00} & \textbf{0.114} & \textbf{Flat+bump} & \textbf{pow} & $\mathbf{1.54^{+0.21}_{-0.22}}$ & \textbf{9.86} & $\mathbf{2.40^{+1.20}_{-0.95}}$  & $\mathbf{45^{+79}_{-26.8}}$ & $\mathbf{8.49^{+0.13}_{-0.14}}$ & (2), (5), (6)\\
	080928 & 1.6919 & 15.5 & 0.179 & SMC-like & pow & $0.29^{+0.03}_{-0.03}$ & $<$9.29 & & & & (1), (5)\\
	\textbf{120119A} & \textbf{1.728} & \textbf{1.77} & \textbf{0.295} & \textbf{SMC} & \textbf{pow} & $\mathbf{1.02^{+0.11}_{-0.11}}$ & \textbf{9.91} & $\mathbf{1.08^{+0.49}_{-0.43}}$ & $\mathbf{43^{+24}_{-14}}$ & $\mathbf{8.6^{+0.14}_{-0.14}}$ & (3), (5), (6)\\
	\hline\hline
	\end{tabular}
	}
	\tablefoot{Columns list (1) the redshift, (2) time at which the SED is extracted, (3) the Galactic extinction, (4) type of extinction curve, (5) whether the GRB afterglow SED is best fitted with a power law or broken power law, (6) the amount of extinction in the V band along the GRB l.o.s in the host galaxy, (7) stellar mass of the GRB host galaxy from \cite{Perley16}, (8) amount of extinction in the host galaxy derived from the Balmer decrement, (9) the SFR derived from H$_{\alpha}$ luminosity, (10) the metallicity, (11) references where these quantities are found. \\
	\textbf{References:} (1) \cite{Zafar11}; (2) \cite{Zafar12}; (3) Zafar et al. (2018, submitted); (4) \cite{Japelj15}; (5) \cite{Perley16}; (6) \cite{Kruehler15}; (7) \cite{Schady12}; (8) \cite{Vergani17} }
	\end{minipage}}
\end{table}

\begin{table}[htbp]
	\resizebox{0.91\textwidth}{!}{\begin{minipage}{\textwidth}
	\centering
	\caption{\label{tab:phot_sample} GRB extinction curves derived by scaling the LMC, SMC and MW extinction curves of \cite{Pei92}. GRBs of the upper part of the table are taken from \cite{Schady12}, GRBs in the lower part are taken from \cite{Japelj15}. The GRBs in bold are the ones in our Golden sample.}
	\renewcommand{\arraystretch}{2}
	\rotatebox{90}{	
	\begin{tabular}{ccccccccccccc}
	\hline\hline
	GRB & z & $\rm T_{SED}$ & $\rm A_{V}^{Gal}$ & ext. curve & pow/bknp & $\rm A_{V}^{grb}$  & log$_{10}$ M$_{\odot}$ & $\rm A_{V}^{lines}$ & SFR$\rm _{H_{\alpha}}$ & $Z$ & References \\
	 & & hr & mag & & & mag & & mag & $M_{\odot}$ yr$^{-1}$ & 12 + log(O/H) & \\
	\hline
	050820A & 2.615 & 2.78 & 0.14 & LMC & bknp & $0.19^{+0.03}_{-0.03}$ & 9.38  & & & & (7), (5)\\
	060418 & 1.49 & 1.39 & 0.69 & SMC & bknp & $0.13^{+0.01}_{-0.02}$ & & & & & (7)\\
	060607A & 3.082 & 2.22 & 0.09 & SMC & pow & $0.08^{+0.04}_{-0.04}$ & $<$9.45 & & & & (7), (5)\\
	060904B & 0.703 & 1.39 & 0.53 & LMC & pow & $0.15^{+0.04}_{-0.04}$ & & & & & (7)\\
	\textbf{061121} & \textbf{1.314} & \textbf{1.67} & \textbf{0.14} & \textbf{LMC} & \textbf{bknp} & $\mathbf{0.54^{+0.06}_{-0.09}}$ & \textbf{10.31} &  $1.66^{+0.60}_{-0.60}$& $44.2^{+19}_{-10}$ & $8.5^{+0.09}_{-0.06}$ & (7), (5), (8)\\
	071112C & 0.823 & 0.36 & 0.36 & SMC & pow & $0.20^{+0.05}_{-0.04}$ & 8.89 & $<$0.2 & & $7.90^{+0.5}_{-0.35}$ & (7), (5), (8)\\
	080210 & 2.641 & 1.39 & 0.26 & LMC & bknp & $0.25^{+0.03}_{-0.03}$ & $<$9.50 & & & & (7), (5)\\
	080319C & 1.95 & 0.14 & 0.08 & SMC & pow & $0.71^{+0.08}_{-0.07}$ & & & & & (7)\\
	080710 & 0.845 & 2.78 & 0.23 & SMC & pow & $0.03^{+0.01}_{-0.01}$ & $<$8.95 & & & & (7), (5)\\
	080804 & 2.204 & 0.42 & 0.05 & LMC & bknp & $0.10^{+0.01}_{-0.01}$ & 9.28 & $1.17^{+1.57}_{-1.08}$  & $15.2^{+41.2}_{-8.7}$ &  & (7), (5), (6) \\
	081121 & 2.512 & 2.77 & 0.16 & SMC & bknp & $0.07^{+0.01}_{-0.01}$ & 9.24 & & & & (7), (5)\\
	081222 & 2.77 & 0.19 & 0.06 & MW & bknp & $0.04^{+0.02}_{-0.01}$ & 9.61 & & & & (7), (5)\\
	090102 & 1.546 & 0.28 & 0.15 & MW & bknp & $0.40^{+0.03}_{-0.03}$ & & & & & (7) \\
	090424 & 0.544 & 0.28 & 0.08 & MW & bknp & $1.14^{+0.09}_{-0.09}$ & 9.62 & $1.42^{+0.54}_{-0.51}$ & $2.88^{+1.81}_{-1.81}$ & $8.88^{+0.12}_{-0.20}$ & (7), (5), (8)\\
	090618 & 0.54 & 0.55 & 0.27 & SMC & bknp & $0.15^{+0.01}_{-0.01}$ & 9.04 & & & & (7), (5)\\
	091018 & 0.971 & 8.33 & 0.09 & SMC & bknp & $0.10^{+0.01}_{-0.01}$ & 9.62 & $0.18^{+1.72}_{-0.18}$ & $1.29^{+3.46}_{-0.32}$ & $8.78^{+0.18}_{-0.19}$ & (7), (5), (6)\\
	091029 & 2.752 & 0.28 & 0.05 & LMC & bknp & $0.06^{+0.03}_{-0.03}$ & $<$9.81 & & & & & (7), (5)\\
	\hline
	\textbf{100814A} & \textbf{1.44} & \textbf{98.4} & \textbf{0.054} & \textbf{SMC} & \textbf{bknp} & $\mathbf{0.20^{+0.03}_{-0.03}}$ & \textbf{9.52} & $\mathbf{0.25^{+0.80}_{-0.25}}$ & $\mathbf{3.2^{+2.9}_{-0.7}}$ & & (4), (5), (6)\\
	100901A & 1.408 & 66.0 & 0.270 & SMC & bknp & $0.29^{+0.03}_{-0.03}$ & & & & & (4)\\
	120815A & 2.358 & 2.06 & 0.320 & SMC & bknp & $0.32^{+0.02}_{-0.02}$ & & $0.18^{+1.05}_{-0.18}$ & $2.3^{+2.7}_{-1}$ & & (4), (5), (6)\\ 
	\hline\hline
	\end{tabular}
	}
	\tablefoot{
	Columns list (1) the redshift, (2) time at which the SED is extracted, (3) the Galactic extinction, (4) type of extinction curve, (5) whether the GRB afterglow SED is best fitted with a power law or broken power law, (6) the amount of extinction in the V band along the GRB l.o.s in the host galaxy, (7) stellar mass of the GRB host galaxy from \cite{Perley16}, (8) amount of extinction in the host galaxy derived from the Balmer decrement, (9) the SFR tracer, (10) the SFR derived from H${_\alpha}$ luminosity, (11) the metallicity, (12) references where these quantities are found.\\
	\textbf{References:} (1) \cite{Zafar11}; (2) \cite{Zafar12}; (3) Zafar et al. (2018, submitted); (4) \cite{Japelj15}; (5) \cite{Perley16}; (6) \cite{Kruehler15}; (7) \cite{Schady12}; (8) \cite{Vergani17}; (8) \cite{Japelj16} }
	\end{minipage}}
\end{table}
\clearpage
\newpage
\twocolumn
\section{SED data for the golden sample}

\subsection{GRBH 061121}
Multi-wavelength data of the host galaxy are available from \cite{Perley15}. It was one of the four galaxies detected in radio from their VLA survey at 3 GHz. 
The galaxy was also observed in the optical (\textit{uBgiz} with the Keck-I/LRIS, \textit{R} with the VLT/FORS2) and near-IR (\textit{J} with the P200/WIRC, \textit{K} with the VLT/ISAAC and 3.6 $\rm \mu$m with Spitzer/IRAC).
\subsection{GRBH 070802}
The galaxy was observed by \cite{Eliasdottir09} with the VLT and detected in the \textit{R} band of the VLT/FORS2 and \textit{K} band of the VLT/ISAAC. \cite{Kruehler11} reported detections with the \textit{i} band of the NTT/EFOSC and \textit{J} band of the VLT/HAWK-I.  Further detections in the NIR are reported by \cite{Perley13} with the HST/WFC3 in F105W and F160W and with Spitzer/IRAC. We decided to keep only the IRAC2 measurement as the IRAC1 measurement is affected by a bright star \citep{Perley13}.
\subsection{GRBH 080605}
The host galaxy is observed by \cite{Kruehler11} with GROND and detected in the \textit{griz} bands. The HST observations (WFC3/F160W) and LIRIS observation (\textit{J}, \textit{Ks}) are reported by \cite{Kruehler12} and IRAC1 magnitude by \cite{Perley16}.
\subsection{GRBH 080607}
We used the photometric data published in \cite{Perley13} from the observations first reported  by \cite{Chen10}: the \textit{gVIK} data from the Keck-I observations (LRIS and NIRC), a  deep \textit{r}-band  measurement from the Magellan/IMACS, and near-IR measurements with HST/WFC3 (F160W)  and Spitzer/IRAC  (IRAC1 and IRAC2). ALMA flux at 345 GHz is reported by \cite{Wang12}. 
\subsection{GRBH 080805}
\cite{Kruehler11} reported photometric observations of the host with the NTT/EFOSC (\textit{V}\textit{R}\textit{i} bands) and the VLT/HAWK-I (\textit{JK} bands). The HST observations (WFC3/F160W) are reported by \cite{Blanchard16} and IRAC1 magnitude by \cite{Perley16}.  We discarded the \textit{i} measurement whose error is twice larger than in the \textit{V,R} bands.
\subsection{GRBH 100814A}
The photometry for the host galaxy is taken from \cite{Nardini14}: we consider the last available \textit{griz} GROND observations where the host dominates the afterglow, 63 days after the burst for \textit{iz} and 145 days after the burst for \textit{gr}. 
\cite{DePasquale15} reported a \textit{J} detection with the Calar Alto 3.5m telescope three years after the burst. We do not consider UVOT data as they are available for only six days after the burst and likely to be contaminated by the GRB afterglow. The Spitzer/IRAC observations were carried out and reported  by \cite{Perley16} and the target was detected in the IRAC1 channel.
\subsection{GRBH 120119A}
Some of the \cite{Morgan14} observations were conducted at a late epoch after the burst. The host galaxy was clearly detected and its emission measured with the Keck/LRIS in the \textit{B}, \textit{R}, and RG850 bands 0.85 year after the burst, and with the HST/WFC3 in the F125W and F160W bands 0.77 year after the burst. The Spitzer/IRAC observations were carried out and reported  by \cite{Perley16} and the target was detected in the IRAC1 channel. 

\clearpage
\newpage
\onecolumn

\section{Best fits with radiative transfer simulations}
\begin{figure}[h]
	\centering
	\includegraphics[width=0.3\columnwidth]{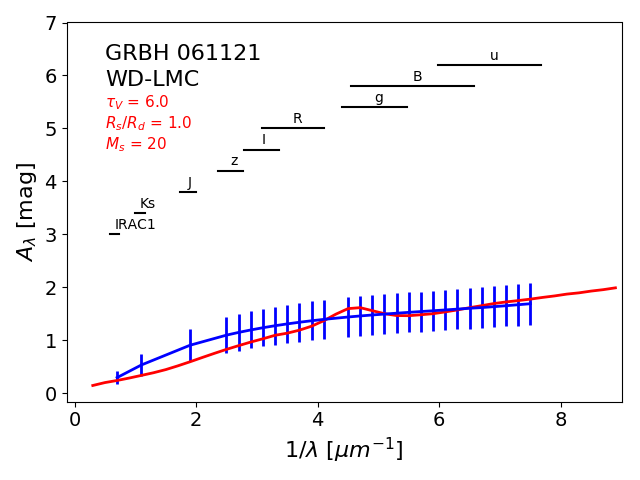}
    	\includegraphics[width=0.3\columnwidth]{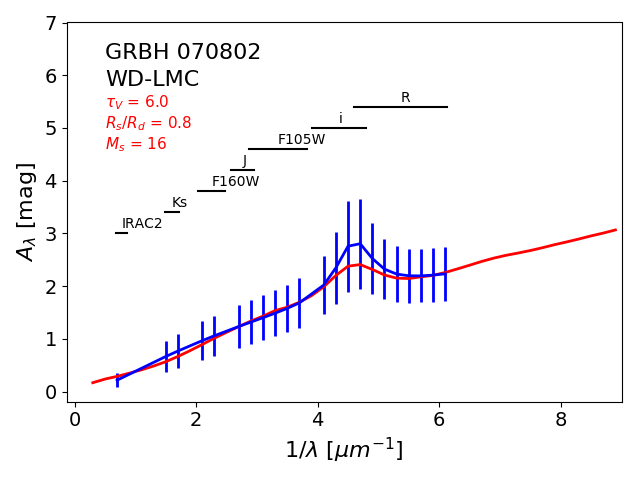}
	\includegraphics[width=0.3\columnwidth]{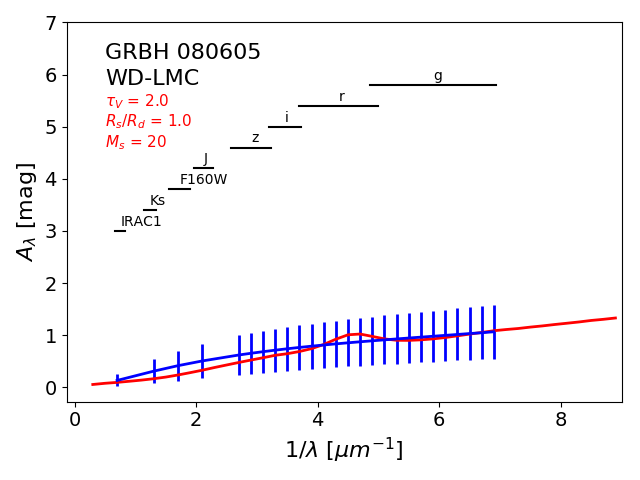}
	\includegraphics[width=0.3\columnwidth]{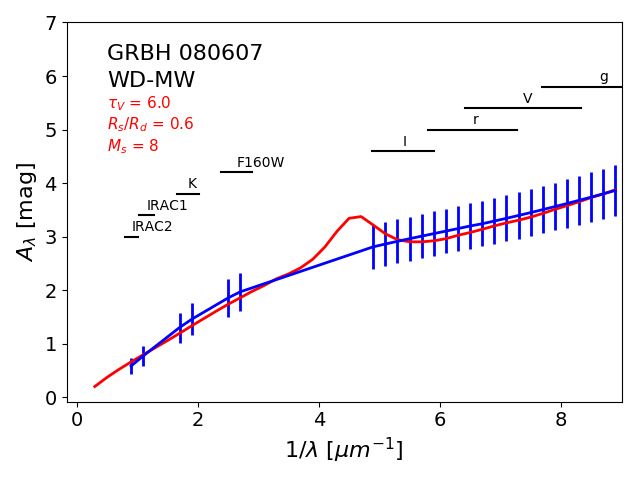}
	\includegraphics[width=0.3\columnwidth]{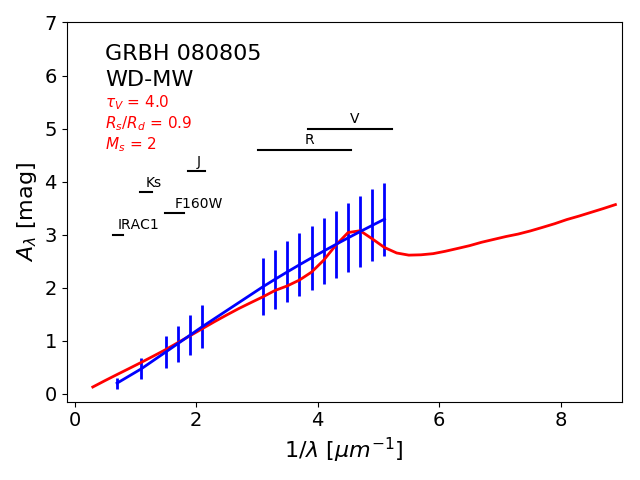}
	\includegraphics[width=0.3\columnwidth]{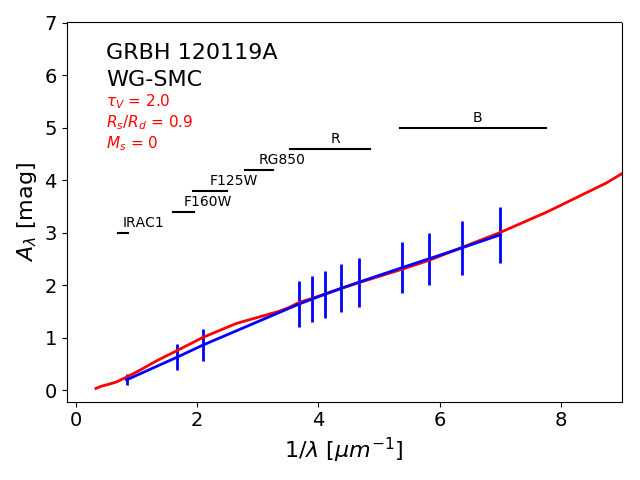}
	\caption[Caption for LOF]{Best fits of the attenuation curve derived with CIGALE with the ones from radiative transfer simulations of \cite{Seon16}.\footnotemark\ The blue errors bars correspond to the attenuation curve and associated uncertainty derived with CIGALE estimated at  wavelengths used in the radiative transfer simulations and covered by observations. The red line corresponds to the attenuation curve from radiative transfer simulations which best fits the CIGALE attenuation curve. The black lines represent the spectral coverage of the host galaxy observations.}
	\label{fig:RT_best_fit}%
\end{figure}

In this appendix, we compare the attenuation curves computed with radiative transfer simulations of \cite{Seon16} best fitting attenuation curves derived with CIGALE in Section \ref{sec:SED_fitting} using the method explained in Section \ref{sec:RT}. Attenuation curves are compared only at  wavelengths used in the radiative transfer simulations and covered by the broad-band filter observations. Only the best fit is reported in Figure \ref{fig:RT_best_fit} with the corresponding values of $\tau_V$, $\frac{R_s}{R_d}$ and $M_s$. In this Figure we also report the rest-frame spectral coverage of each host galaxy observation.
\footnotetext{We note that fitting GRBH 061121 with a WG-SMC dust grain model results in a better fit without UV bump but it does not change qualitatively the description of the ISM structure as the best values for parameters are: $\tau_V$=6.0,  $\frac{R_s}{R_d}$=1.0 and $M_s$=16.}
\end{appendix}

\end{document}